\documentclass[acmsmall,nonacm]{acmart}

\newcommand{\mytablesize}{\scriptsize}
\usepackage{url}
\usepackage{multirow}
\usepackage{siunitx}
\usepackage{hyperref}
\usepackage{graphicx}
\usepackage{subfig}
\usepackage[labelfont=bf]{caption}
\newcolumntype{P}[1]{>{\centering\arraybackslash}p{#1}}
\usepackage{enumitem}
\setlist{  
  listparindent=\parindent,
  parsep=0pt,
}

\usepackage{xcolor}
\definecolor{myblue}{rgb}{0.333, 0.447, 0.878}
\definecolor{myred}{rgb}{0.816, 0.286, 0.247}

\def\Posbar#1{~#1 {\color{myred}\rule{#1cm}{6pt}} }
\def\Negbar#1{#1 {\color{myblue}\rule{-#1cm}{6pt}} }

\makeatletter
\newcommand\footnoteref[1]{\protected@xdef\@thefnmark{\ref{#1}}\@footnotemark}
\makeatother

\usepackage{blindtext}

\renewcommand\footnotetextcopyrightpermission[1]{} 
\AtBeginDocument{%
  \providecommand\BibTeX{{%
    \normalfont B\kern-0.5em{\scshape i\kern-0.25em b}\kern-0.8em\TeX}}}

\setcopyright{acmcopyright}
\copyrightyear{2021}
\acmYear{2021}

\makeatletter
\let\@authorsaddresses\@empty
\makeatother

\acmConference[]{}{}{}

\begin{document}

\title{Nowcasting Gentrification Using Airbnb Data}

\author{Shomik Jain}
\affiliation{
  \institution{Center for AI in Society's Student Branch, University of Southern California}
  \country{United States}
}

\author{Davide Proserpio}
\affiliation{
  \institution{Marshall School of Business, University of Southern California}
  \country{United States}
}

\author{Giovanni Quattrone}
\affiliation{
  \institution{Middlesex University, United Kingdom -}
  \institution{University of Turin}
  \country{Italy}
}

\author{Daniele Quercia}
\affiliation{
  \institution{King's College, United Kingdom -}
  \institution{Nokia Bell Labs}
  \country{United Kingdom}
}


\begin{abstract}
There is a rumbling debate over the impact of gentrification: presumed gentrifiers have been the target of protests and attacks in some cities, while they have been welcome as generators of new jobs and taxes in others. Census data fails to measure neighborhood change in real-time since it is usually updated every ten years. This work shows that Airbnb data can be used to quantify and track neighborhood changes. Specifically, we consider both structured data (e.g., number of listings, number of reviews, listing information) and unstructured data (e.g., user-generated reviews processed with natural language processing and machine learning algorithms) for three major cities, New York City (US), Los Angeles (US), and Greater London (UK). We find that Airbnb data (especially its unstructured part) appears to nowcast neighborhood gentrification, measured as changes in housing affordability and demographics. Overall, our results suggest that user-generated data from online platforms can be used to create socioeconomic indices to complement traditional measures that are less granular, not in real-time, and more costly to obtain.
\end{abstract}


\maketitle

\section{Introduction}\label{sec:intro}
Gentrification is a revitalization process characterized by physical and socioeconomic changes in urban neighborhoods. These changes usually involve in-movers that are affluent, educated, or younger compared to out-movers that are poor, uneducated, or older~\citep{freeman_gentrification, lees_gentrification, zuk_gentrification}. Disadvantaged neighborhoods are especially vulnerable to gentrification because their residents resemble these out-movers and because these neighborhoods have experienced disinvestment from the public or private sectors~\citep{freeman_gentrification}. Indeed, over 20\% of disadvantaged neighborhoods across the US have gentrified since 2000~\citep{maciag_20percent}. Of particular concern in these neighborhoods is gentrification-induced displacement, a phenomenon in which out-movers are forced to move for reasons beyond their control~\citep{zuk_gentrification}. To mitigate these negative effects of gentrification, governments and municipalities have tried many strategies such as rent-control and public investment~\citep{levy_mitigate}.

Implementing gentrification policies requires first identifying gentrifying neighborhoods. Since gentrification is associated with socioeconomic changes, governments have measured gentrification using demographic data from public agencies such as the US Census Bureau and UK Office of National Statistics (ONS)~\citep{zuk_gentrification}. However, these agencies rely on survey-based methods for obtaining demographic data, which pose several problems. First, government surveys are expensive: The 2020 US Census will cost over \$15 billion~\citep{us_census_cost}, and the 2021 UK Census will cost at least \$1 billion~\citep{uk_census_cost}. This amounts to \$50-\$100 just to survey a single household on average. Second, government data are quickly outdated because they represent a fixed point in time and have a delayed-release. The main Census in both the US and UK occurs every 10 years. In addition, the US Census Bureau reports 5-year estimates of demographic data obtained through the American Community Surveys (ACS), and the UK ONS reports the Indices of Multiple Deprivation (IMD) every 4 years. Due to these problems, both the US and UK governments have expressed concerns about the future of survey-based methods~\citep{us_census_concerns, uk_census_concerns}. 

The nowcasting of socioeconomic indices related to gentrification would constitute a major improvement over the status quo of outdated government data. With nowcasted information, policymakers could make data-driven decisions to address the negative effects of gentrification in the present. For these reasons, there has been a growing interest in using alternative sources of data to measure and nowcast important urban and economic outcomes, as we shall see in Section~\ref{sec:related}.~\citep{glaeser_yelp} use Yelp data to quantify neighborhood changes,~\citep{naik_streetscore} propose ``Streetscore'', a scene-understanding algorithm that predicts the perceived safety of a streetscape using Google Maps Street View data and, finally,~\cite{glaeser_streetscore} show that they can predict the median income of residents in New York City from Google Maps images using a computer vision model. It is worth pointing out that prior work has focused on nowcasting gentrification -- as opposed to forecasting it -- mainly because alternative data sources have grown substantially only in recent years, making the validation of any forecasting model very challenging.

In this paper, we use user-generated data from Airbnb -- a popular peer-to-peer short term rental platform -- to nowcast gentrification. Specifically, our work contributes to the growing scientific and public debate about short-term rentals and their urban impact~\citep{quattrone_spatial, wachsmuth2018airbnb, yrigoy2016impact}. We use a combination of structured data (e.g., listing information) and unstructured data (e.g., the textual content of reviews) processed with a variety of machine learning techniques. Given gentrification is most prevalent in large cities~\citep{zuk_gentrification}, this work focuses on two major US cities, New York City and Los Angeles, and one major European city, London. We nowcast gentrification measured as changes in socioeconomic variables between two temporal windows, 2013--2017 and 1998--2002, from Airbnb data in 2013--2017, and make two main contributions: 

\begin{enumerate}
    \item We mine a variety of data sources and profile neighborhoods in terms of gentrification scores and Airbnb features (Section~\ref{sec:data}). For each neighborhood, we construct a gentrification score based on changes in socioeconomic measures of age, education, housing affordability, and income (Section~\ref{sec:score}). We then collect both structured Airbnb data (e.g., number of listings, number of reviews, listing information) and unstructured Airbnb data (e.g., user-generated reviews), which we analyze and process using  machine learning algorithms such as Latent Dirichlet Allocation (LDA)~\citep{lda} and representation learning~\citep{bengio2013representation} (Section~\ref{sec:airbnb_data}).
    
    \item We then study the ability of Airbnb to nowcast gentrification (Section~\ref{sec:results}). We start by showing that there is a high correlation between gentrification and Airbnb data (Section~\ref{sec:results_corr}). We also find that unstructured data processed with machine learning tools tend to have more explanatory power (Section~\ref{sec:results_uns}), suggesting that unstructured data can capture aspects of gentrification beyond those captured by structured data alone. Given these strong correlations, we develop models using Airbnb data that show the potential to nowcasting gentrification with user-generated information (Section~\ref{sec:results_predicting}). 
\end{enumerate}

\section{Related Work}\label{sec:related}

Recent years have seen a rapid growth in online platforms and user-generated information. Social media platforms such as Facebook, Twitter, and Instagram allow users to freely share content and opinions with both friends and strangers; online review platforms such as TripAdvisor and Yelp allow anyone to write reviews about any kind of service, from hotels to restaurants to even health-care; and sharing economy platforms such as Airbnb allow people to review the homes and neighborhoods of strangers. The value of user-generated information is growing as the size of the data increases, as more people use these digital platforms, and as more people rely on such data to make decisions. For example, a growing body of research provides evidence suggesting that online reviews and ratings can affect firms' sales and revenues~\citep{chevalier2006effect,luca2016reviews}. Turning to social media, researchers have exploited user-generated content to predict consumer preferences and behavior~\citep{zhang2013predicting}, and brand perception~\citep{liu2018visual,culotta2016mining}, among other applications.

Additionally, there has been a growing interest in using alternative sources of data to measure and predict important urban and economic outcomes. This is for two reasons. First, user-generated content is readily available for free, it is available at any geographical granularity (zipcode, city, or state level) and, more importantly, it is available at high frequency and in real-time. Second, in the last few years, there have been tremendous improvements in computation power and algorithms to manage and process large amounts of structured and unstructured data.  Within this literature, we find a diverse set of papers with different goals. In terms of macroeconomic indices, Antenucci et al.~\citep{antenucci_twitter} and Proserpio et al.~\citep{proserpio_twitter} use Twitter data to predict labor market outcomes such as the unemployment rate. With respect to cities and more granular measures, Naik et al.~\citep{naik_streetscore} propose ``Streetscore'', a scene-understanding computer vision algorithm that measures the perceived safety of neighborhoods using Google Maps Street View data. Their subsequent work~\citep{glaeser_streetscore} shows that Google Maps images can predict the median income of residents in New York City neighborhoods. Cranshaw et al.~\citep{cranshaw2012CSCW} and Venerandi et al.~\citep{Venerandi2015CSCW} also use Foursquare data to quantify urban trends, and Hristova et al.~\citep{flickr_frontiers} use Flickr data to quantify cultural capital and predict changes in house prices. Finally, the work closest to ours is the work of Glaeser et al.~\citep{glaeser_yelp}, who show that the number of businesses listed on Yelp is highly correlated to changes in socioeconomic variables related to gentrification including age, education, and housing. The main difference between~\citep{glaeser_yelp} and our work is that we use a combination of structured and unstructured data to predict gentrification.

Moreover, several papers have explored the characteristics of the neighborhoods in which Airbnb enters and the effects of this platform on economic activity. Quattrone et al.~\citep{quattrone_spatial, quattrone_sharing} analyze the spatial penetration of Airbnb in the US and UK and show that Airbnb listings tend to appear more often in neighborhoods occupied by the ``talented and creative'' classes, which resemble the in-movers of the gentrification process. Their subsequent work~\cite{quattrone_location} analyzes Airbnb reviews and shows how this unstructured data contains important nuances that are neighborhood dependent. Furthermore, Basuroy et al.~\citep{basuroy_sleeping} show that increases in Airbnb listings in Texas zipcodes are associated with increases in economic activity in these zipcodes.

While previous work considers unstructured image data to measure urban outcomes, in this work, we focus on textual information and the process of gentrification. Our focus on text is driven by the hypothesis that user-generated data related to short-term rentals could contain latent valuable information about neighborhoods and their economic conditions, thus helping cities and municipalities better measure and understand the process of gentrification.

\section{Profiling Neighborhoods}\label{sec:data}

To nowcast gentrification, we create a gentrification score that captures changes in neighborhood socioeconomic conditions and create features from Airbnb data to predict such changes. We define neighborhoods as zipcodes in the US and as wards in the UK because these represent the most granular administrative divisions for which governments provide socioeconomic data. Zipcodes and wards are defined based on population sizes: a zipcode contains 8,000 people on average whereas a ward contains about 5,500 people. Thus, the geographical sizes of each vary based on population density. In total, there are about 200 zipcodes in New York City, 130 zipcodes in Los Angeles, and 630 wards in Greater London. 

\subsection{Gentrification Score}\label{sec:score}
We construct an overall gentrification score for each neighborhood (zipcode or ward) based on changes in four socioeconomic measures: age, education, housing affordability, and income (Table~\ref{tab:socioecon}). Since these measures are not published in the same year and come from a variety of data sources, we group them by two temporal windows, 1998--2002 and 2013--2017. Then, between the two windows, a gentrification score can be computed.

Our gentrification score definition is similar to prior work that quantifies gentrification by aggregating changes in socioeconomic conditions between two time periods~\citep{freeman_gentrification, lees_gentrification, zuk_gentrification, bousquet_gentrification}. Historically, governments have also measured gentrification using similar demographic data from public agencies such as the US Census Bureau and UK Office of National Statistics (ONS)~\citep{bousquet_gentrification}. We specifically use the socioeconomic variables of age, education, housing affordability, and income because public agencies in both the US and UK collect these variables. While race has also been found to be associated with gentrification~\citep{zuk_gentrification}, this variable is only available in the US. Therefore, we decided to omit this variable in our main analysis for consistency. However, in Table 7 in the Appendix, we show that including race in the gentrification score of US cities leads to similar results.

\begin{table}[t!]
\mytablesize
\caption{Socioeconomic Measures for US Zipcodes and UK Wards.}
\begin{tabular}{p{1.2cm}p{1.5cm}p{4cm}p{2cm}p{2cm}}
\toprule
\multirow{2}{*}{Country} & \multirow{2}{*}{Measure} & \multirow{2}{*}{Definition} & Data Source & Data Source \\
& & & 1998--2002 & 2013--2017 \\
\midrule
\multirow{5.5}{*}{US} & Age & Percent aged between 25 and 34 & \multirow{5.5}{2cm}{2000 Decennial Census~\citep{census_data}} & \multirow{5.5}{2cm}{2013-2017 \\ Census American Community Surveys~\citep{census_data}} \\
\addlinespace
& Education & Percent with a bachelors degree \\
\addlinespace
& Housing & Median gross rent \\
\addlinespace
& Income & Median household income \\
\addlinespace
\midrule
\multirow{6}{*}{UK} & Age & Percent aged between 25 and 34 & 2002 ONS~\citep{ons_age} & 2014 ONS~\citep{ons_age} \\
\addlinespace
& Education & Attainment and skills in the population & \multirow{5}{2cm}{2000$^\dagger$ ONS \\ Indices of Multiple Deprivation \citep{ons_imd}} & \multirow{5}{2cm}{2015$^\dagger$ \& 2019$^\dagger$ ONS Indices of Multiple Deprivation \citep{ons_imd}} \\
\addlinespace
& \multirow{2}{*}{Housing} & Lack of physical and financial \\
& & accessibility to housing \\
\addlinespace
& Income & Percent not deprived from low income \\
\bottomrule
\multicolumn{5}{l}{\textit{$^\dagger$The ONS collects data for the Indices of Multiple Deprivation two years prior to release~\citep{ons_imd_delay}.}}
\end{tabular}
\label{tab:socioecon}
\end{table}

First, we aggregate our socioeconomic measures to create a neighborhood index for each zipcode or ward. To standardize each measure, we consider the percentile within a given city instead of the raw value. We also use percentiles because the raw Indices of Multiple Deprivation (IMD) values are not comparable across different years~\citep{ons_imd_delay}. Then, we construct a neighborhood index for each neighborhood and temporal window $\mathit{tw}$: 
\begin{equation}
    \text{Neighborhood Index}_{\mathit{tw}} =  \frac{1}{4} \times \left( \text{Age}_{\mathit{tw}} + \text{Education}_{\mathit{tw}} + \text{Housing}_{\mathit{tw}} + \text{Income}_{\mathit{tw}} \right).
\end{equation}
A lower neighborhood index indicates that a neighborhood is more disadvantaged on the basis of more old, uneducated, or poor residents as well as cheaper housing. Similar to prior work~\citep{freeman_gentrification, lees_gentrification, zuk_gentrification}, we define (and limit our analysis to) disadvantaged neighborhoods as those having a neighborhood index in the bottom $50^{th}$ percentile in the first time window (1998--2002). In doing so, we are left with 83 disadvantaged zipcodes in New York City, 68 disadvantaged zipcodes in Los Angeles, and 230 disadvantaged wards in London.

After standardizing each neighborhood index using its percentile, we define the gentrification score for disadvantaged neighborhoods in line with previous definitions~\citep{freeman_gentrification, lees_gentrification, zuk_gentrification}:
\begin{equation}
    \text{Gentrification Score}_{\mathit{tw}} = \left( \text{Neighborhood Index}_{\mathit{tw}}  - \text{Neighborhood Index}_{\mathit{tw}-1} \right),
\end{equation}
where $\mathit{tw}$ is 2013--2017 and $\mathit{tw}-1$ is 1998--2002. 
A higher gentrification score indicates that a neighborhood has experienced more gentrification on the basis of an influx of young, educated, or wealthy residents as well as decreased housing affordability.

In defining gentrification, we focus solely on disadvantaged neighborhoods for two reasons. First, non-disadvantaged neighborhoods experience significantly less change (t-test, $p<0.05$) in the neighborhood index between 1998--2002 and 2013--2017. Second, and more importantly, the concept of gentrification is usually discussed only in the context of disadvantaged neighborhoods. Even though affluent neighborhoods may experience some change in socioeconomic measures, this is typically not considered to be gentrification~\cite{zuk_gentrification}. 

\begin{figure}[t!]
    \centering
    \includegraphics[width=\linewidth]{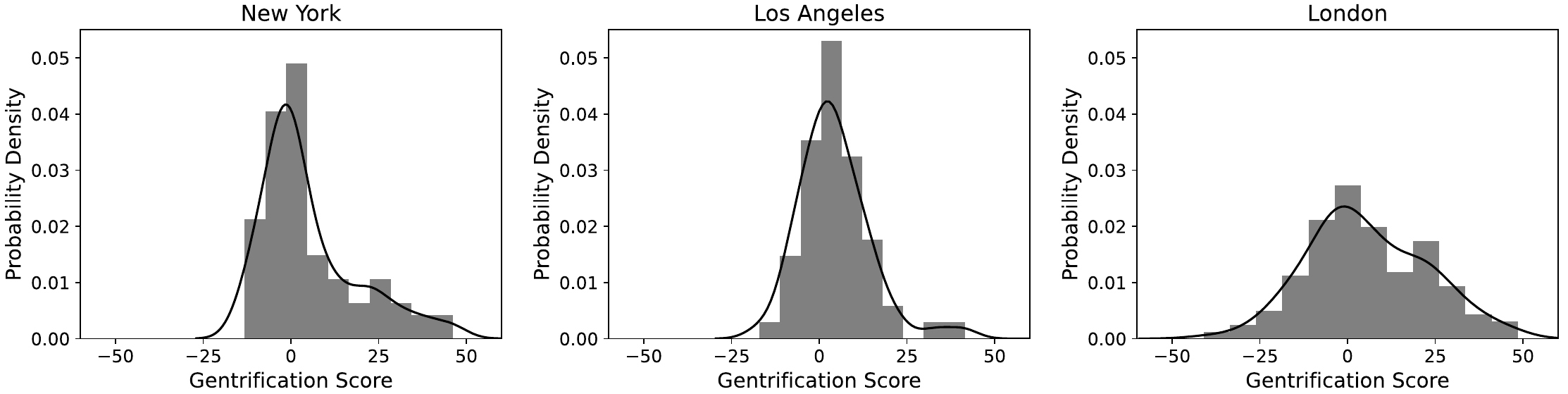}
    \caption{Distribution of the Gentrification Score.}
    \label{fig:dist_score}
\end{figure}

Figure~\ref{fig:dist_score} shows the distribution of the gentrification score among disadvantaged neighborhoods. We observe that each distribution is centered around zero, indicating that neighborhoods experience no change on average in their gentrification score. However, there is significant variation in the gentrification score across all cities. Table~\ref{table:prediction} reports summary statistics for the gentrification score, neighborhood index, and each socioeconomic measure. 



To further validate our gentrification score, we performed a sensitivity analysis to test the extent to which our results depend on its definition. First, we found a strong correlation ($r=0.65$) with an existing gentrification score for the city of Los Angeles~\citep{bousquet_gentrification}. Second, we performed a cross-correlation test among the four socioeconomic measures composing it (age, education, income, rent), and found an average cross-correlation of $r=0.40$ (see Figure~\ref{fig:socioeconomic_correlation} in the Appendix). The strong cross-correlation scores also help justify our decision to equally weight the four socioeconomic measures, despite the possibility that each measure could have different levels of importance in each neighborhood. Third, to test the robustness of our results, we replicated our analysis for each socioeconomic variable individually and obtained similar results (see Tables 9-12 in the Appendix).

\begin{table}[t!] \centering \mytablesize
\caption{Summary Statistics for Gentrification Measures in Disadvantaged Neighborhoods.}
\begin{tabular}{p{3cm}p{1.6cm}
S[table-number-alignment = center, group-digits=true, table-format= 3.3, group-separator={}, round-mode=places, round-precision=2]
S[table-number-alignment = center, group-digits=true, table-format= 3.3, group-separator={}, round-mode=places, round-precision=2]
S[table-number-alignment = center, group-digits=true, table-format= 3.3, group-separator={}, round-mode=places, round-precision=2]
S[table-number-alignment = center, group-digits=true, table-format= 3.3, group-separator={}, round-mode=places, round-precision=2]
S[table-number-alignment = center, group-digits=true, table-format= 3.3, group-separator={}, round-mode=places, round-precision=2]
S[table-number-alignment = center, group-digits=true, table-format= 3.3, group-separator={}, round-mode=places, round-precision=2]
} 
\toprule
\multirow{2}{3cm}{Measure} & \multirow{2}{1.6cm}{Time Period} &\multicolumn{2}{c}{New York} &\multicolumn{2}{c}{Los Angeles} &\multicolumn{2}{c}{London}\\ 
\cmidrule(lr){3-4} \cmidrule(lr){5-6} \cmidrule(lr){7-8}
& & \multicolumn{1}{c}{Median} & \multicolumn{1}{c}{SD}
& \multicolumn{1}{c}{Median} & \multicolumn{1}{c}{SD}
& \multicolumn{1}{c}{Median} & \multicolumn{1}{c}{SD} \\
\midrule
\multirow{2}{3cm}{Age} & 1998-2002 & 44.34 & 18.401 & 68.947 & 21.591 & 59.062 & 21.927 \\
 & 2013-2017 & 55.189 & 20.984 & 70.0 & 22.453 & 66.631 & 23.395\\
\addlinespace
\multirow{2}{3cm}{Education} & 1998-2002 & 32.547 & 15.255 & 24.561 & 16.619 & 24.733 & 18.085\\
 & 2013-2017 & 36.321 & 18.684 & 27.544 & 20.676 & 32.303 & 18.627\\
\addlinespace
\multirow{2}{3cm}{Income} & 1998-2002 & 31.132 & 15.143 & 14.386 & 13.544 & 22.068 & 17.028\\
 &  2013-2017 & 32.075 & 15.73 & 19.123 & 14.724 & 25.267 & 17.683\\
\addlinespace
\multirow{2}{3cm}{Rent} &  1998-2002 & 32.075 & 13.821 & 14.912 & 13.49 & 26.599 & 20.166\\
 &  2013-2017 & 34.434 & 16.339 & 19.825 & 17.807 & 28.465 & 22.699\\
\addlinespace
\midrule
\multirow{2}{3cm}{Neighborhood Index} & 1998-2002 & 30.66 & 11.164 & 20.614 & 13.416 & 25.586 & 14.647 \\ 
 & 2013-2017 & 33.491 & 17.691 & 24.035 & 19.703 & 30.384 & 20.645\\
\addlinespace
\midrule
Gentrification Score & 1998-2017 &  0.236 & 13.809 & 2.807 & 14.286 & 3.198 & 16.914\\
\addlinespace
\midrule
\multicolumn{2}{l}{Disadvantaged Neighborhoods} & \multicolumn{2}{c}{79} & \multicolumn{2}{c}{58} & \multicolumn{2}{c}{186} \\
\bottomrule
\end{tabular}
\label{table:prediction}
\end{table}

\begin{table}[h!] \centering \mytablesize
\caption{Summary Statistics for Airbnb Data in 2013--2017 from Disadvantaged Neighborhoods.}
\begin{tabular}{p{4.6cm}
S[table-number-alignment = center, group-digits=true, table-format= 4.2, group-separator={}, round-mode=places, round-precision=2]
S[table-number-alignment = center, group-digits=true, table-format= 4.2, group-separator={}, round-mode=places, round-precision=2]
S[table-number-alignment = center, group-digits=true, table-format= 4.2, group-separator={}, round-mode=places, round-precision=2]
S[table-number-alignment = center, group-digits=true, table-format= 4.2, group-separator={}, round-mode=places, round-precision=2]
S[table-number-alignment = center, group-digits=true, table-format= 4.2, group-separator={}, round-mode=places, round-precision=2]
S[table-number-alignment = center, group-digits=true, table-format= 4.2, group-separator={}, round-mode=places, round-precision=2]
} 
\toprule
\multirow{2}{*}{Airbnb Data Features} &\multicolumn{2}{c}{New York} &\multicolumn{2}{c}{Los Angeles} &\multicolumn{2}{c}{London}\\
\cmidrule(lr){2-3} \cmidrule(lr){4-5} \cmidrule(lr){6-7}
& \multicolumn{1}{c}{Median} & \multicolumn{1}{c}{SD}
& \multicolumn{1}{c}{Median} & \multicolumn{1}{c}{SD}
& \multicolumn{1}{c}{Median} & \multicolumn{1}{c}{SD} \\
\midrule
\textit{Structured Data Features} \\
\hspace{1em} \#~Listings &  135.000 & 1050.265 & 165.0 & 566.251 & 28.0 & 46.192\\
\addlinespace
\hspace{1em} \#~Reviews & 1997.0 & 16203.768 & 2709.0 & 13768.619 & 461.0 & 1271.529\\
\addlinespace
\hspace{1em} Price (USD) &  81.718 & 22.795 & 87.617 & 31.139 & 64.076 & 29.906\\
\addlinespace
\hspace{1em} \#~Bedrooms &  1.154 & 0.095 & 1.155 & 0.183 & 1.344 & 0.251\\
\addlinespace
\hspace{1em} Star-Rating &  4.618 & 0.111 & 4.687 & 0.144 & 4.69 & 0.141\\
\addlinespace
\hspace{1em} Location Star-Rating &  8.99 & 0.359 & 9.06 & 0.535 & 9.286 & 0.339\\
\addlinespace
\midrule
\textit{Unstructured Data Features} \\
\hspace{1em} Review Length (words) &  23.256 & 2.532 & 21.251 & 2.139 & 22.806 & 3.612\\
\addlinespace
\hspace{1em} Location Words (\%) &  19.605 & 2.383 & 16.925 & 3.404 & 19.636 & 3.806\\
\addlinespace
\hspace{1em} Sentiment &  0.798 & 0.037 & 0.809 & 0.043 & 0.852 & 0.042\\
\addlinespace
\hspace{1em} Sentiment in Location Reviews &  0.856 & 0.032 & 0.847 & 0.036 & 0.887 & 0.036\\
\addlinespace
\hspace{1em} LDA Components (Altogether) &  0.17 & 0.125 & 0.166 & 0.118 & 0.158 & 0.112\\
\addlinespace
\hspace{1em} Doc2Vec Components (Altogether) &  -0.006 & 0.097 & 0.009 & 0.092 & -0.006 & 0.092\\
\addlinespace
\midrule
Disadvantaged Neighborhoods & \multicolumn{2}{c}{79} & \multicolumn{2}{c}{58} & \multicolumn{2}{c}{186} \\
\bottomrule
\end{tabular}
\label{table:features}
\end{table}

\begin{table}[h!] 
\centering \mytablesize
\caption{LDA Topics.} 
\scalebox{0.9}{
\begin{tabular}{
P{4cm}
P{4cm}
P{4cm}}
\toprule
\textbf{New York} & \textbf{Los Angeles} & \textbf{London} \\
\midrule
\multicolumn{3}{c}{\textbf{Topic: Check-In}} \\
\addlinespace
\multirow{1}{4cm}{arrive, host, day, cancel, reserve, post, automatic, check, flexible, late, smooth, process, early, flight, key} & \multirow{1}{4cm}{arrive, day, host, reserve, cancel, post, automatic, check, late, airbnb, instruct, time, book, text, didn't, key}  &  \multirow{1}{4cm}{day, arrive, host, room, bed, bathroom, night, kitchen, check, reserve, cancel, post, shower, automatic, work}\\
\\
\\
\midrule
\multicolumn{3}{c}{\textbf{Topic: Listing Characteristics}} \\
\addlinespace
\multirow{1}{4cm}{room, apartment, bed, bathroom, night, stay, clean, place, good, kitchen, time, didn't, bedroom, check, sleep} & \multirow{1}{4cm}{room, bed, park, apartment, clean, place, bathroom, stay, nice, night, good, kitchen, work, bedroom, towel} & \multirow{1}{4cm}{great, location, nice, good, clean, place, stay, room, apartment, host, easy,, flat, communication, close, check} \\
\\
\\
\midrule
\multicolumn{3}{c}{\textbf{Topic: Location}} \\
\addlinespace
\multirow{1}{4cm}{apartment, great, restaurant, location, stay, love, walk, place, shop, perfect, view, bar, close, park, subway, area} & \multirow{1}{4cm}{walk, beach, location, santa, monica, great, close, apartment, place, distance, restaurant, park, stay, shop, hollywood} & \multirow{1}{4cm}{great, flat, location, restaurant, love, stay, london, apartment, shop, close, area, walk, park, perfect, recommend}\\
\\
\\
\midrule
\multicolumn{3}{c}{\textbf{Topic: Stay/Host}} \\
\addlinespace
\multirow{1}{4cm}{great, stay, place, location, host, clean, apartment, recommend, nice, definitely, help, comfort, friend, perfect, time} & \multirow{1}{4cm}{great, stay, place, location, clean, host, nice, apartment, recommend, definitely, easy, help, comfort, good, perfect} & \multirow{1}{4cm}{stay, place, great, host, love, london, recommend, clean, help, location, house, friend, room, definitely, comfort} \\
\\
\\
\midrule
\textbf{Topic: Public Transportation} & \textbf{Topic: Stay/Host} & \textbf{Topic: Public Transportation} \\
\addlinespace
\multirow{1}{4cm}{subway, place, close, walk, great, location, nice, good, stay, apartment, minute, clean, station, time, manhattan} & \multirow{1}{4cm}{stay, love, house, place, host, time, beautiful, comfort, friend, feel, wonder, perfect, recommend, amazing, la} & \multirow{1}{4cm}{station, walk, minute, london, tube, bus, place, close, min, 5, 10, train, house, stay, underground} \\
\\
\\
\bottomrule
\end{tabular}}
\label{table:lda_topics}
\end{table}

\subsection{Airbnb Features}\label{sec:airbnb_data}
Airbnb launched in 2008 as a peer-to-peer platform for short-term rental accommodations. Hosts can list their properties for rent on the platform, and guests can book these properties for a few days to multiple weeks. Airbnb has experienced exponential growth during the past decade, starting in 2008 with just 100 listings in San Francisco to now offering over 6 million listings in 192 countries. More than 500 million guests have stayed with Airbnb to date. 

From the Airbnb website, we collected the complete set of listings and reviews data in New York City, Los Angeles, and London. We chose these cities because they have a history of gentrification~\citep{lees_gentrification,maciag_20percent} as well as the most Airbnb data in the US and UK. Specifically, we consider Airbnb data for the 5-year period 2013--2017 so that it aligns with the most recent period for which we have socioeconomic data. The collected data amounts to over 180K listings and 3M reviews in New York City, Los Angeles, and Greater London. In our analysis, we focus on disadvantaged neighborhoods and exclude those with less than 5 listings (10th percentile), leaving us with 49,765 listings and 768,450 reviews in 79 New York City zipcodes; 22,908 listings and 477,758 reviews in 58 Los Angeles zipcodes; and 8,617 listings and 181,037 reviews in 186 London wards. 

We create two types of features from the Airbnb data (Table~\ref{table:features}): those obtained from structured data (e.g., number of listings, number of reviews, listing information) and those obtained from unstructured data (e.g., user-generated reviews). All features are aggregated at the neighborhood level (zipcode or ward) over the 5-year period for which they were collected, which helps reduce potential noise in Airbnb data and also avoids the use of any Personally Identifiable Information (PII)\footnote{While the Airbnb website publicly displays some PII in listing details and reviews, in our analyses we do not rely on any individual user data, but aggregate data at the neighborhood level.}.

\paragraph{\normalfont{\textbf{Structured Data Features}}} Structured data features consist of the following information about listings, aggregated at the neighborhood level over 2013--2017: 
\begin{itemize}\setlength\itemsep{0.5em}
    \item \textit{\#~Listings} and \textit{\#~Reviews}: the total number of listings and reviews.
    \item \textit{Price}: the average price per listing.
    \item \textit{\#~Bedrooms}: the average number of bedrooms available for rent per listing.  
    \item \textit{Star-Rating} and \textit{Location Star-Rating}: the average overall and location star-rating per listing. We exclude star-ratings for other topics such as cleanliness, accuracy, value, communication, and check-in because we did not find these features to be correlated ($|r|<0.15$) with gentrification.
\end{itemize}

\paragraph{\normalfont{\textbf{Unstructured Data Features}}} Unstructured data features are constructed from the review text using Natural Language Processing (NLP) techniques. We first preprocess\footnote{We calculate the sentiment feature without preprocessing review text because VADER accounts for unorthodox text such as punctuation, slang, and acronyms.} the review text to remove punctuation and commonly used words, and stem each remaining word to its root form. Then, we compute the following features, aggregated over all English reviews in a neighborhood over 2013--2017. 

\begin{itemize}\setlength\itemsep{0.5em}
    \item \textit{Review Length}: The average number of words in each review.
    \item \textit{Location Words}: The average percentage of location-related words in each review. Quattrone et al.~\citep{quattrone_location} analyze Airbnb reviews and create a vocabulary of commonly-used social and business words in Airbnb reviews. Social words are those focusing on the interaction between guests and hosts (e.g., words like ``sharing'', ``talking'', ``chatting'', ``conversation''), whereas business words are those focusing on the business transaction between guests and hosts (e.g., words concerning the property, its location, or the professional conduct of the host). We use their shared and public dictionary to analyze different word categories and find that the frequency of location-related business words (e.g., location, place, neighborhood, area) is the most relevant for gentrification.\footnote{Location Words Dictionary: \url{https://figshare.com/s/991c8677e3e9ce013774}}
    \item \textit{Sentiment}: The average sentiment of each review. We calculate sentiment on a scale of $-1$ (most extreme negative) to $+1$ (most extreme positive) using the Valence Aware Dictionary and Sentiment Reasoner (VADER)~\citep{vader_sent}. VADER is a lexicon and rule-based sentiment analysis tool specifically attuned to sentiments expressed in social media. 
    \item \textit{Sentiment in Location Reviews}: The average sentiment of location-related reviews. We define location-related reviews as those which 10\% of the words are location words. 
    \item \textit{LDA Components}: The average presence of each LDA topical component in each review. LDA~\citep{lda} is an unsupervised topic extraction model that uses word frequencies to group text samples into latent topical components. First, LDA determines the associated words for a given number of latent topics. Then, for each text sample, it outputs topic scores to represent the probabilities that the sample corresponds to each topical component. Following standard practice, we determine the optimal number of topics (five in our case) using the perplexity score. We report the top-15 words for each topic in Table~\ref{table:lda_topics}. Based on these words, we determine the latent topics in all cities to be related to four common subjects: ``check-in'', ``listing characteristics'', ``location'', and ``stay/host''.\footnote{Additionally, New York City and London have a public transportation topic that does not appear in the city of Los Angeles where public transportation is neither well developed or widely used.} 
    \item \textit{Doc2Vec Components}: The average Doc2Vec vector coordinates for each review. Doc2Vec~\citep{d2v} is an unsupervised representation learning method~\citep{bengio2013representation} that maps text to vectors in an $n$-dimensional space. We use Doc2Vec with 25 dimensions and obtain a vector representation of each review. The output vectors of Doc2Vec preserve semantic information about the input text; in particular, reviews that have similar word frequencies are closer in the $n$-dimensional vector space.
\end{itemize}


\section{Nowcasting Gentrification Using Airbnb Data}\label{sec:results}
We now test the extent to which we can nowcast gentrification from Airbnb data in disadvantaged neighborhoods. First, we examine the correlations between the gentrification score and both structured and unstructured Airbnb data features (Section~\ref{sec:results_corr}). Next, we discuss the insights that we can obtain from unstructured data (Section~\ref{sec:results_uns}). Finally, we nowcast the gentrification score using both in-sample linear regression and out-of-sample random forest regression (Section~\ref{sec:results_predicting}). 

\begin{figure}[t!]
\centering
\subfloat[][Gentrification Score between 1998--2002 and 2013--2017]{\includegraphics[width=\linewidth]{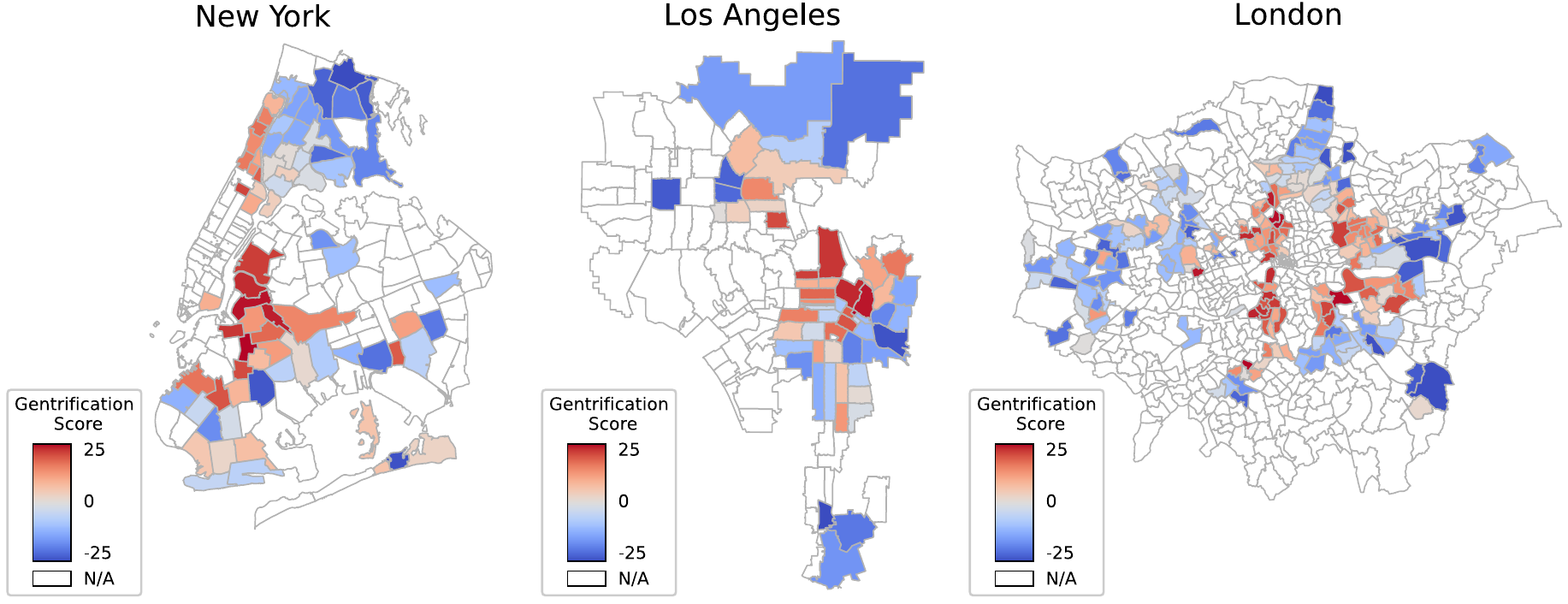}}
\newline
\subfloat[][Number of Airbnb Listings in 2013--2017]{\includegraphics[width=\linewidth]{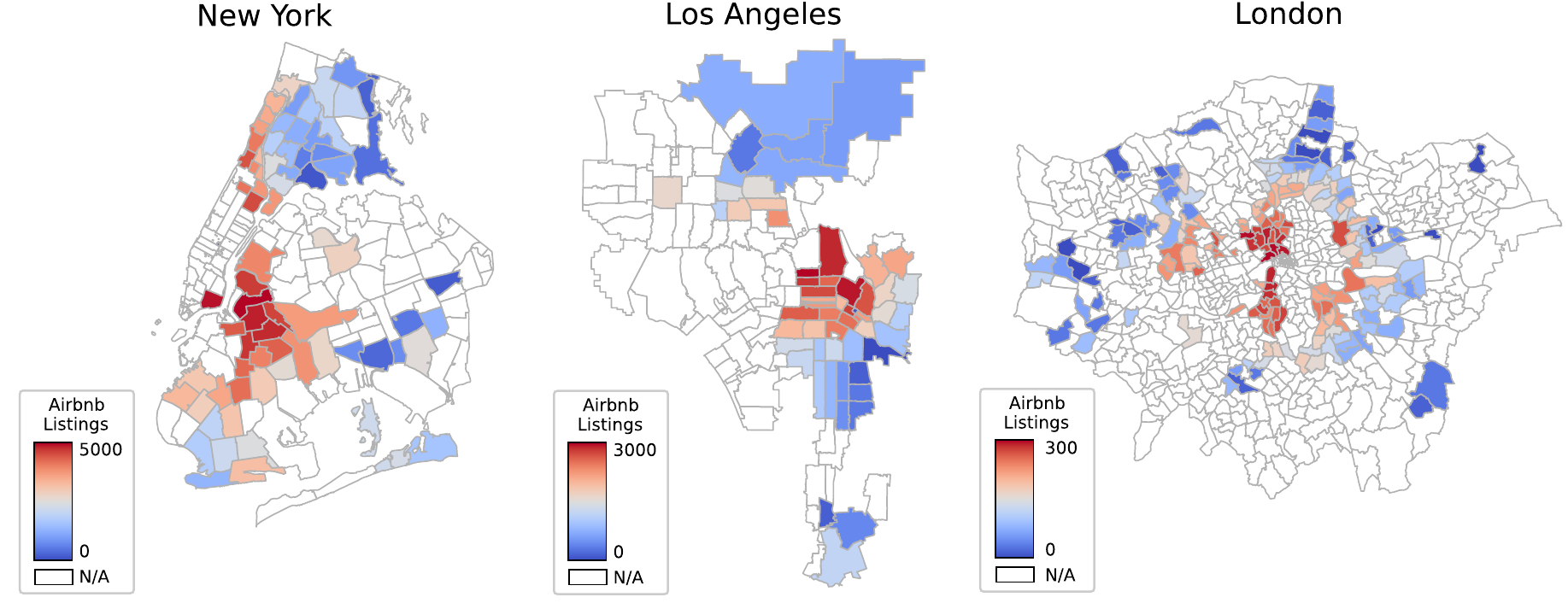}}
\caption{Comparison of the Gentrification Score and Number of Airbnb Listings.}
\label{fig:map}
\end{figure}

\subsection{Correlation between Airbnb Data and Gentrification}\label{sec:results_corr}
We start by analyzing the correlation between gentrification and Airbnb data across disadvantaged neighborhoods. Table~\ref{table:correlation} reports the linear correlation coefficients ($r$) between each Airbnb data feature and the gentrification score. 

\paragraph{\normalfont{\textbf{Structured Data Features}}} Among structured data features, we find high and positive correlation ($p<0.05$) in all cities for the number of listings ($r\geq0.40$), the number of reviews ($r\geq0.37$), and the listing price ($r\geq0.30$). In other words, gentrifying neighborhoods have more Airbnb listings (as Figure~\ref{fig:map} confirms) as well as more reviews and higher listing prices. The overall star-rating does not have a significant correlation ($p>0.1$) in any city. However, the location star-rating (which rates the Airbnb listings' location) has a significant ($p<0.05$) positive correlation in New York City ($r=0.28$) and London ($r=0.30$). The number of bedrooms available for rent has different effects across cities, being negatively correlated to gentrification in Los Angeles and positively correlated in London; we argue this is likely due to the feature capturing the unique geography and housing availability of each city.

\paragraph{\normalfont{\textbf{Unstructured Data Features}}} Most unstructured data features are highly correlated with the gentrification score. For all cities, the location words feature has a correlation of $0.41$ ($p<0.01$) or higher, and the review length has a correlation of $0.27$ ($p<0.01$) or higher. These positive correlations suggest that users write longer reviews that contain more location-related words in gentrifying neighborhoods. In addition, we find that the sentiment in reviews mentioning location has higher correlation than the overall sentiment for all cities. This suggests that, if a review talks more positively about location, then it is more likely that the corresponding neighborhood is gentrifying. Finally, we find that most of the LDA topics are highly correlated with the gentrification score, with the location topic having the highest correlation in all three cities ($r\geq0.43$, $p<0.01$). Similarly, most of the Doc2Vec components are also correlated with the gentrification score, with the highest correlated component having a correlation (in absolute value) above 0.43 ($p<0.01$) in all three cities. Next, we proceed to discuss the interpretation of the Doc2Vec and LDA features.

\begin{table}[ht]
    \mytablesize
    \tabcolsep 1pt
    \centering
    \caption{Linear Correlation ($r$) between Gentrification Score and Airbnb Data.} 
	\begin{tabular}{p{5cm} rl @{\hskip 10pt} rl @{\hskip 10pt} rl}
        \toprule
        Airbnb Data Variables & 
        \multicolumn{2}{c}{New York} & 
        \multicolumn{2}{c}{Los Angeles} & 
        \multicolumn{2}{c}{London}\\ 
        \midrule
        \textit{Structured Data Features} \\
        \hspace{1em} \#~Listings            & *** & \Posbar{0.682} & *** & \Posbar{0.397} & *** & \Posbar{0.547} \\
        \hspace{1em} \#~Reviews             & *** & \Posbar{0.637} & *** & \Posbar{0.370} & *** & \Posbar{0.473} \\
        \hspace{1em} Price                  & *** & \Posbar{0.431} & *** & \Posbar{0.298} & *** & \Posbar{0.353} \\
        \hspace{1em} \#~Bedrooms            &     & \Posbar{0.027} & **  & \Negbar{-0.279}& **  & \Posbar{0.145} \\
        \hspace{1em} Star-Rating            &     & \Posbar{0.056} &     & \Negbar{-0.214}&     & \Posbar{0.081} \\
        \hspace{1em} Location Star-Rating   & **  & \Posbar{0.277} &     & \Posbar{0.204} & *** & \Posbar{0.300}\\
        \midrule
        \textit{Unstructured Data Features} \\
        \hspace{1em} Review Length                              & *** & \Posbar{0.519} & **  & \Posbar{0.271} & *** & \Posbar{0.344} \\
        \hspace{1em} Location Words                             & *** & \Posbar{0.412} & *** & \Posbar{0.446} & *** & \Posbar{0.427} \\
        \hspace{1em} Sentiment                                  & *** & \Posbar{0.391} &     & \Posbar{0.108} & *** & \Posbar{0.209} \\
        \hspace{1em} Sentiment in Location Reviews              & *** & \Posbar{0.469} & *   & \Posbar{0.235} & *** & \Posbar{0.231} \\
        \hspace{1em} LDA Component (Check-In)                   &     & \Negbar{-0.024}&     & \Negbar{-0.012}& **  & \Negbar{-0.169} \\
        \hspace{1em} LDA Component (Listing Characteristics)    & *** & \Negbar{-0.332}&     & \Negbar{-0.187}&     & \Posbar{0.083} \\
        \hspace{1em} LDA Component (Location)                   & *** & \Posbar{0.436} & *** & \Posbar{0.464} & *** & \Posbar{0.423} \\
        \hspace{1em} LDA Component (Stay/Host)                  & *** & \Negbar{-0.320}& **  & \Negbar{-0.324}& *** & \Negbar{-0.287} \\
        \hspace{1em} LDA Component (Public Transportation)      &     & \Posbar{0.009} &     & N/A            &     & \Posbar{0.069} \\
        \hspace{1em} Doc2Vec Component (top correlated comp.)                      & *** & \Negbar{-0.655}& *** & \Posbar{0.476} & *** & \Negbar{-0.437} \\
        \midrule
        Disadvantaged Neighborhoods & \multicolumn{2}{c}{79} & \multicolumn{2}{c}{58} & \multicolumn{2}{c}{186} \\
        \bottomrule
        \multicolumn{7}{r}{\textit{Significance levels: }{$^{*}$p$<$0.1; $^{**}$p$<$0.05; $^{***}$p$<$0.01}} \\
	\end{tabular}
    \label{table:correlation}
\end{table}

\begin{figure}[h!]
\centering
\subfloat[][New York]{\includegraphics[width=0.3\linewidth]{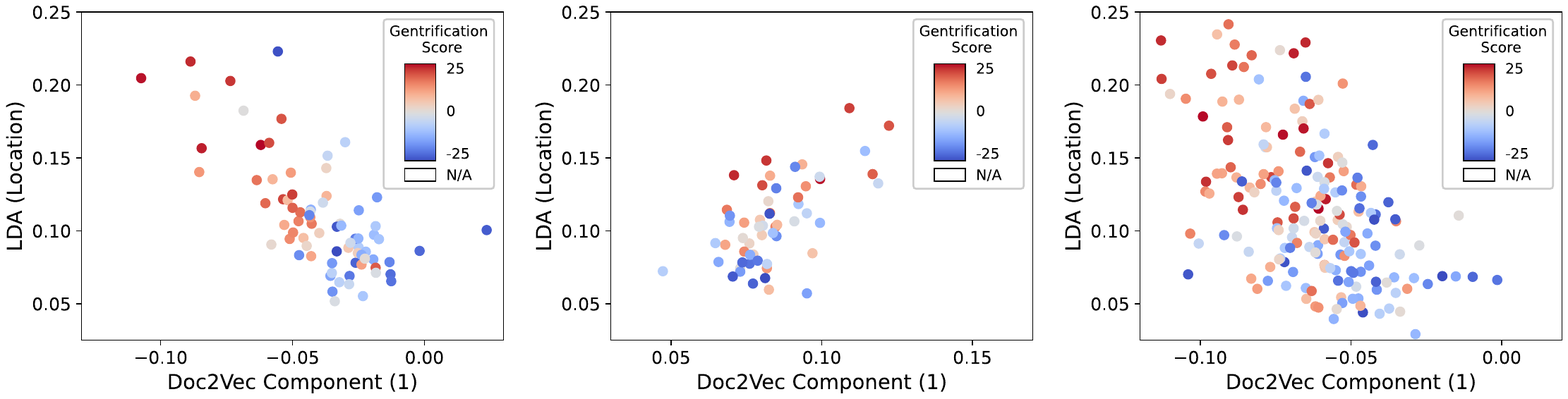}}
\hspace{0.03\linewidth}
\subfloat[][Los Angeles]{\includegraphics[width=0.3\linewidth]{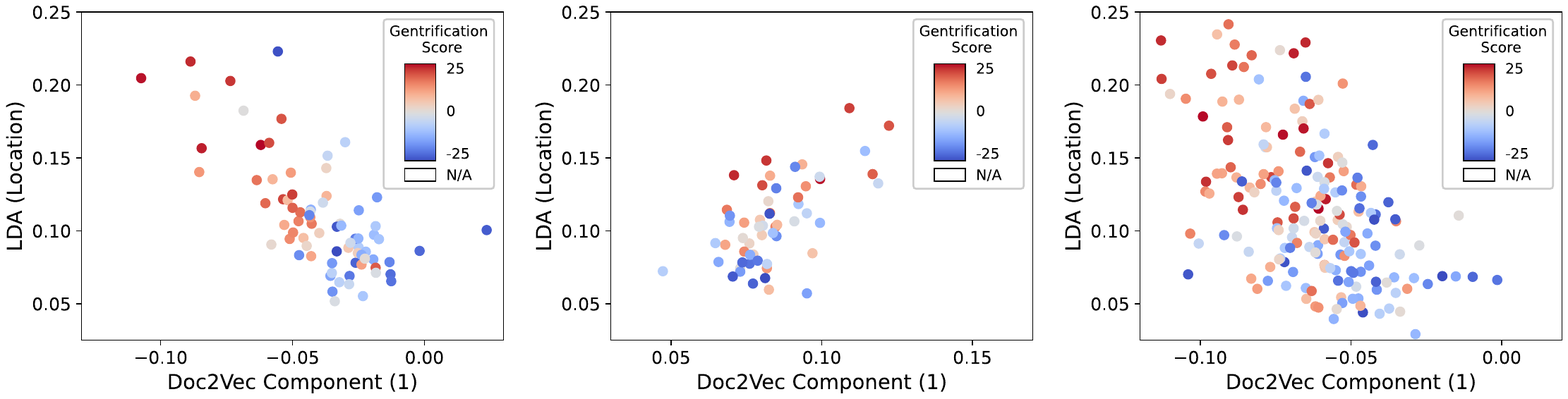}}
\hspace{0.03\linewidth}
\subfloat[][London]{\includegraphics[width=0.3\linewidth]{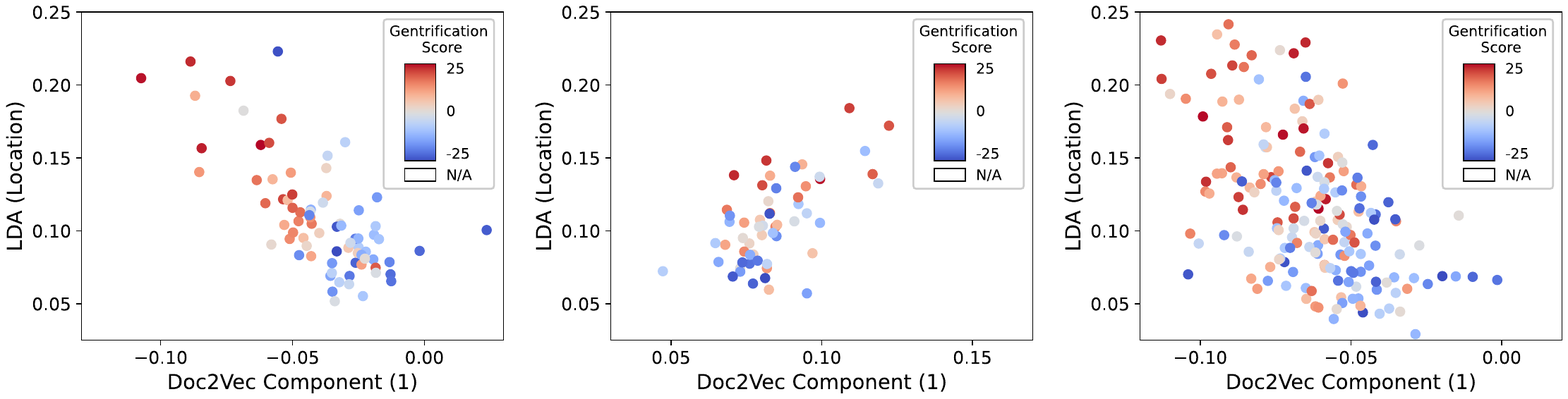}}
\caption{The relationship between LDA and Doc2Vec components and the gentrification score. Each point represents a neighborhood, and points are colored with their level of gentrification, from lower (blue) to higher (red). 
}
\label{fig:corr_lda_Doc2Vec}
\end{figure}

\begin{figure}[t!]
    \centering
    \includegraphics[width=0.55\linewidth]{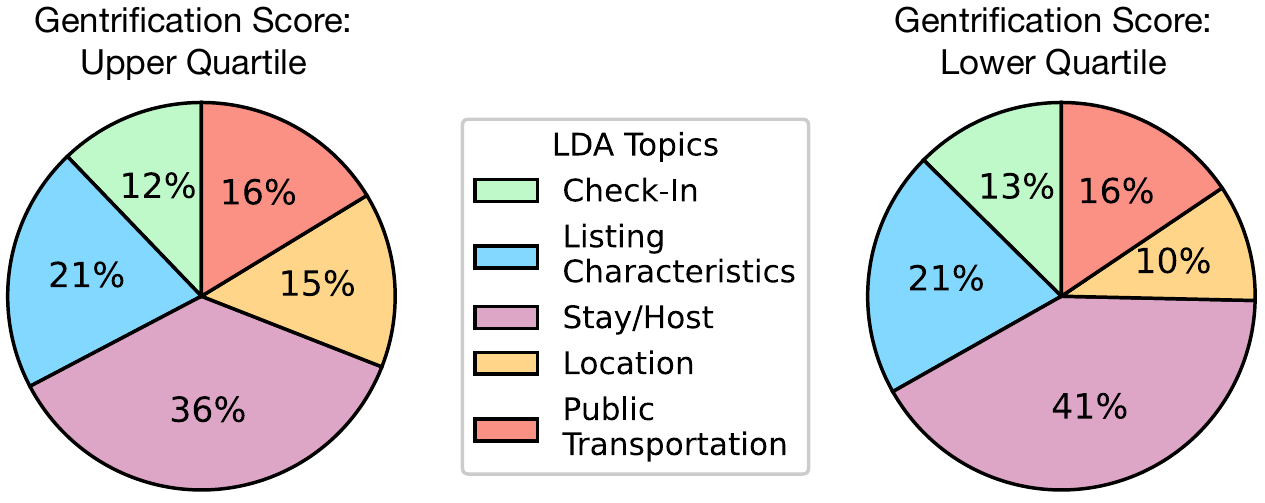}
    \caption{Distribution of LDA topics for neighborhoods in the upper quartile of the gentrification score and neighborhoods in the lower quartile of the gentrification score.}
    \label{fig:lda_pie}
\end{figure}

\begin{figure}[t!]
    \centering
    \includegraphics[width=0.75\linewidth]{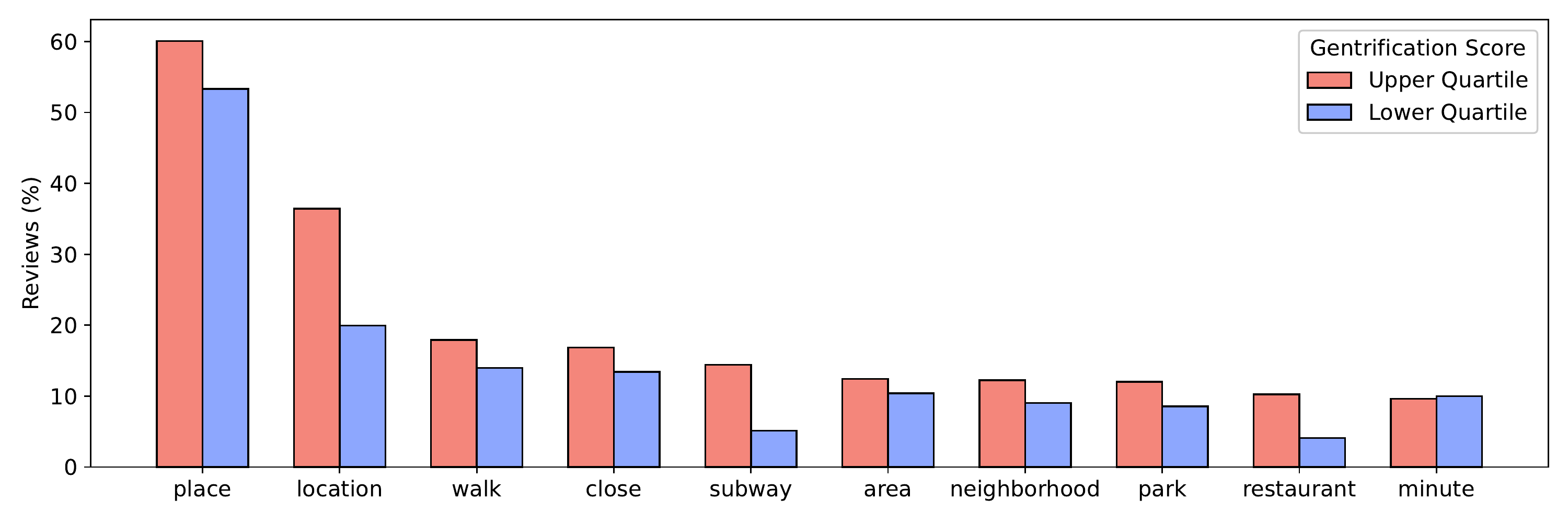}
    \caption{Comparison of location words usage between neighborhoods in the upper quartile of the gentrification score and neighborhoods in the lower quartile of the gentrification score.}
    \label{fig:location_words}
\end{figure}

\subsection{Insights from Unstructured Airbnb Data}
\label{sec:results_uns}


To analyze and interpret our unstructured data features, we select the LDA and Doc2Vec components that have the highest correlation with the gentrification score. For each city, Figure~\ref{fig:corr_lda_Doc2Vec} plots each neighborhood (as a point) according to its values for the two components. The neighborhoods are colored with their level of gentrification score, from lower (blue) to higher (red). We observe that the resulting arrangement highly corresponds with the coloring of the points; for example in Panel (c), London wards with a higher gentrification score cluster in the top-left. The ability to partition neighborhoods by gentrification score using these two unstructured features suggests that strong markers of gentrification exist in the actual content of Airbnb reviews for all of our three cities. 

To further interpret these Doc2Vec and LDA latent components, we measure the correlation of these components with the location words feature, and find high correlation for both of them ($|r|\geq0.49$, $p<0.01$).\footnote{The interested reader can find the whole cross-correlation table for all considered features in the Appendix.} This suggests that these components are capturing the location information contained in the reviews. 

Next, we look at the distribution of topics in reviews for neighborhoods in the upper quartile of the gentrification score and reviews in the lower quartile of the gentrification score. We find that the location topic is present in a significantly larger (t-test, $p<0.01$) proportion of reviews, again confirming that latent location information contained in the text of the reviews is crucial to identifying gentrifying neighborhoods.

Finally, to understand which location words matter the most, we compare the frequency of these words for gentrifying and non-gentrifying neighborhoods. To do so, we first find the top-10 location words in all reviews. We then compute their frequency in reviews from neighborhoods in the upper quartile of the gentrification score and in reviews from neighborhoods in the lower quartile of the gentrification score. Figure~\ref{fig:location_words} reports these results. We observe that location words do indeed appear significantly more frequently (t-test, $p<0.01$) in reviews from upper quartile neighborhoods. Moreover, we find that the most used location words describe the neighborhood (e.g., walk, subway, restaurant, or park) suggesting that Airbnb guests not only describe the listings in which they stay but also the surrounding neighborhood.  



\begin{figure}[t!]
\centering
\subfloat[][In-Sample Linear Regression]{\includegraphics[width=0.45\linewidth]{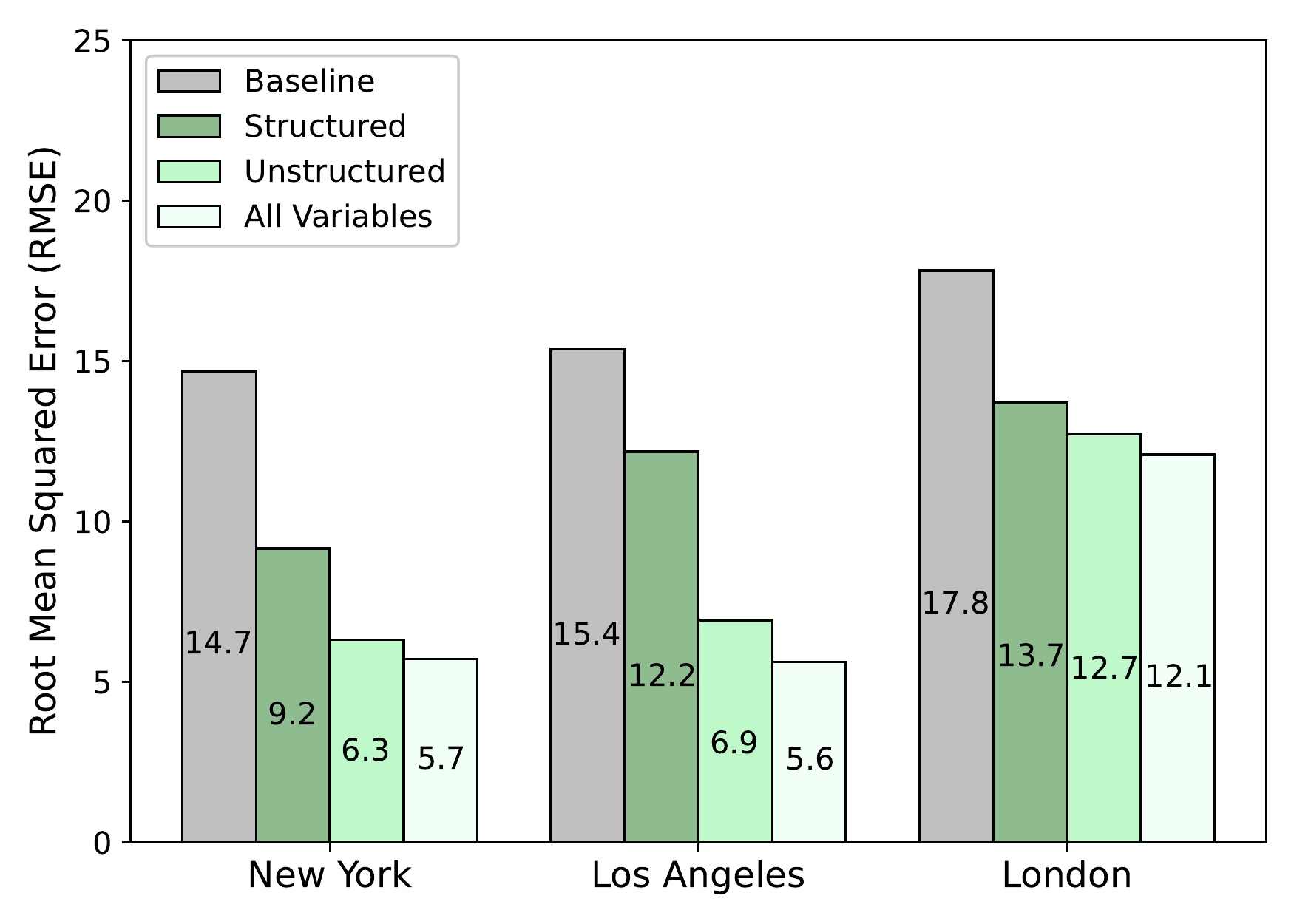}}
\hspace{0.03\linewidth}
\subfloat[][Out-of-Sample Random Forest Regression]{\includegraphics[width=0.45\linewidth]{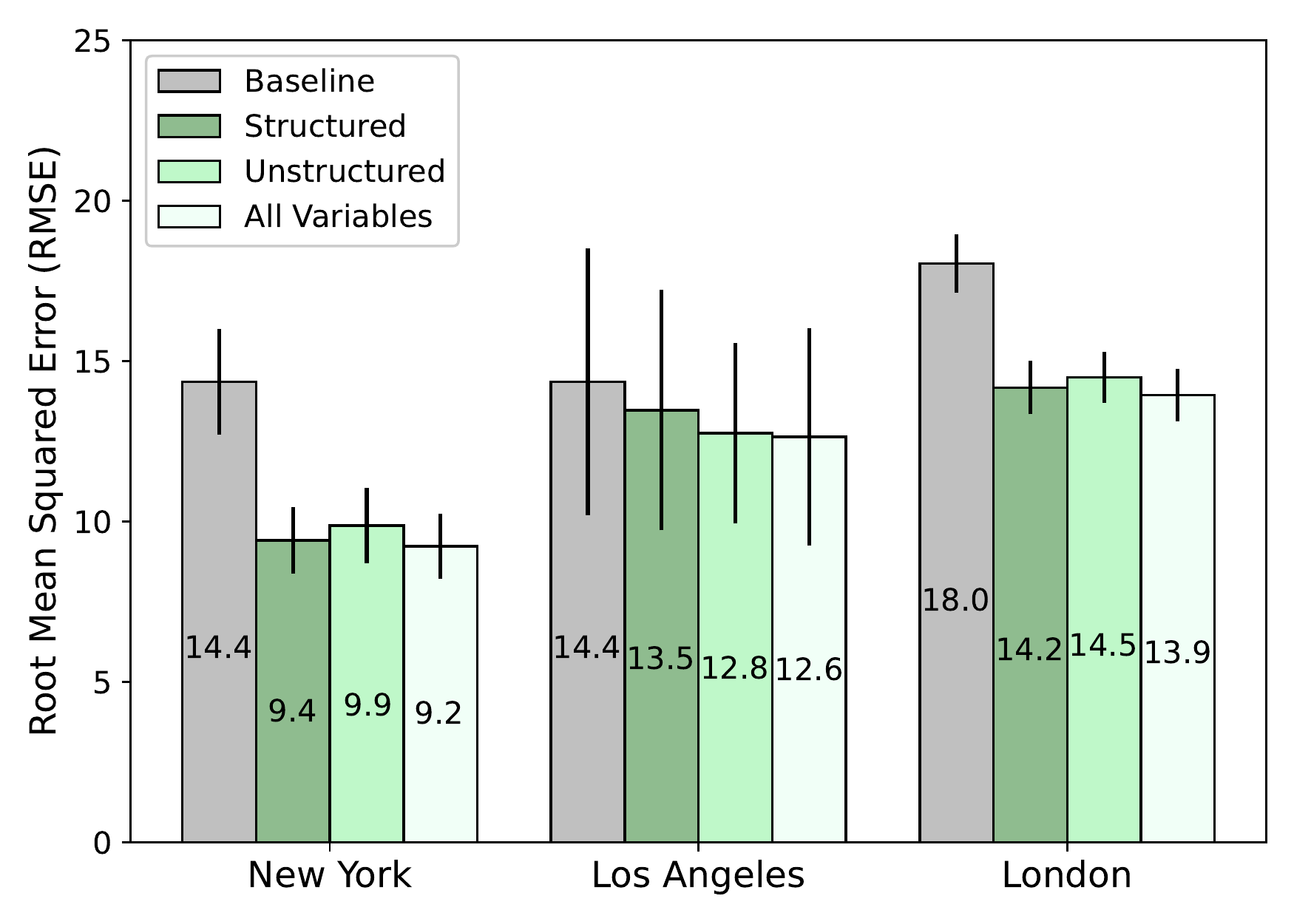}}
\caption{Results from in-sample linear regression (left) and out-of-sample random forest regression (right). We compare our results to a baseline model in which we predict no gentrification. Random forests results represent the average over 100 simulations with 50\%-50\% train-test split, and error bars represent the standard deviation.}
\label{fig:regression}
\end{figure}

\begin{table}[t!] \centering \mytablesize
\caption{Feature Importance from Random Forest Regression.}
\begin{tabular}{P{0.15cm}
P{2.7cm}
S[table-number-alignment = center, group-digits=true, table-format= 2.2, group-separator={}, table-space-text-post = \%, round-mode=places, round-precision=2]
P{3cm}
S[table-number-alignment = center, group-digits=true, table-format= 2.2, group-separator={}, table-space-text-post = \%, round-mode=places, round-precision=2]
P{2.7cm}
S[table-number-alignment = center, group-digits=true, table-format= 2.2, group-separator={}, table-space-text-post = \%, round-mode=places, round-precision=2]
} 
\toprule
& \multicolumn{2}{c}{New York} &
\multicolumn{2}{c}{Los Angeles} & \multicolumn{2}{c}{London} \\ 
\cmidrule(lr){2-3} \cmidrule(lr){4-5} \cmidrule(lr){6-7}
& \multicolumn{1}{c}{Feature} & \multicolumn{1}{c}{MDI} &
\multicolumn{1}{c}{Feature} & \multicolumn{1}{c}{MDI} &
\multicolumn{1}{c}{Feature} & \multicolumn{1}{c}{MDI} \\
\midrule
1 & \#~Listings & 12.4782\% & \#~Listings & 10.7878\% & \#~Listings & 12.1161\%  \\
2 & \#~Reviews & 10.9079\%  & \#~Reviews & 9.6727\% & \#~Reviews & 10.2318\%  \\
3 & Doc2Vec (1) & 8.5230\% & Location Star-Rating & 6.9735\% & Doc2Vec (1) & 6.2435\% \\
4 & Doc2Vec (2) & 7.8366\%  & LDA (Location) & 6.4906\% & Location Words & 6.0784\% \\
5 & Doc2Vec (3) & 7.4698\%  & Doc2Vec (1) & 6.3508\% & Price & 1.6435\% \\
\bottomrule
\multicolumn{7}{l}{\textit{Note: Features are ranked by their Mean Decrease Impurity (MDI) which calculates feature importance as the percentage of times a feature }} \\
\multicolumn{7}{l}{\textit{is used to split a node, weighted by the number of samples it splits.}} \\
\end{tabular}
\label{table:rf_features}
\end{table}

\subsection{Nowcasting Gentrification}
\label{sec:results_predicting}
In this section, we use regression models to predict the gentrification score with Airbnb data. We start with a simple linear regression to predict in-sample gentrification. Then, we turn to random forest regression for out-of-sample predictions.\footnote{We opt for random forest regression due to high multicollinearity among our predictors and the fact that linear regression tends to suffer from it. See Figure~\ref{fig:inter_correlations} in the Appendix for the cross-correlations among our Airbnb features.} Throughout this section, we compare our results across models using only structured features, only unstructured features, and a combination of both. Further, we compare these results to a baseline model that predicts no gentrification (gentrification score $=0$). We use this baseline for two reasons. First, on average, neighborhoods in our dataset do not experience gentrification, i.e., the difference in socioeconomic variables between the time windows that we consider is zero (see Figure~\ref{fig:dist_score}); second, this baseline is equivalent to a scenario in which we would only have access to socioeconomic data from the early period (1998--2002), which is often the case given the large gaps and delays in government data release. 



\paragraph{\normalfont{\textbf{In-Sample Predictions}}} To predict in-sample gentrification scores, we use a linear regression model estimated using ordinary least squares (OLS). We report the Root Mean Square Errors (RMSE) for the four specifications (baseline, structured, unstructured, and all features) in Panel (a) of Figure~\ref{fig:regression}. In-sample linear regression yields RMSE ranging from 5.63 to 12.09 using all features and across all cities. These results significantly outperform the baseline model, which obtains RMSE from 14.70 to 17.82. Interestingly, we achieve similar performance for New York and Los Angeles but much lower performance for London. The smaller geographical size of wards compared to zipcodes may explain this difference. For example, wards contain only 28 listings on average, while zipcodes contain about 150 listings (Table~\ref{table:features}), suggesting the importance of having a sufficient number of listings and reviews for the geographic unit of analysis. 


The specifications using only structured or unstructured data features also outperform the baseline, with unstructured features (6.32 $\leq$ RMSE $\leq$ 12.73) obtaining better results than structured features (9.16 $\leq$ RMSE $\leq$ 13.72) across all cities. However, the specification using all features outperforms those using structured or unstructured features alone. This result suggests that structured and unstructured data are complimentary in nature, and that unstructured data can capture aspects of gentrification that structured data alone cannot. This also confirms our hypothesis that the text of the reviews written by Airbnb guests contain important latent information that can help cities and municipalities better measure and understand the process of gentrification.  

\paragraph{\normalfont{\textbf{Out-of-Sample Predictions}}} To predict out-of-sample gentrification scores, we use a random forest regression. In order to test the robustness of our model, we use default hyper-parameters, and average results for random 50\%--50\% train-test splits across 100 simulations. Through these simulations, we aim to show the generalizability of our model. In other words, we want to test whether we can predict the gentrification score by knowing the underlying socioeconomic measures only in certain neighborhoods. Ideally, if these types of analyses achieve good results (low errors), then they could be used to complement or substitute for traditional analyses based on expensive and outdated survey data.

Random forest regression yields out-of-sample RMSE ranging from 9.23 to 13.95 when using all features and across all cities. Similar to the in-sample analysis, these errors are lower than those obtained from the baseline model, which has RMSE ranging from 13.36 to 18.05. However, we find that the results vary substantially across cities. We achieve the best performance in New York City, where the 9.23 out-of-sample RMSE is similar to the 5.72 in-sample RMSE from linear regression. London also performs comparatively well to its in-sample results (13.95 out-of-sample RMSE; 12.09 in-sample RMSE). As Figure~\ref{fig:regression}(b) shows, the standard deviation of RMSE across simulations is also small for these cities (<1.0). On the other hand, Los Angeles performs much worse compared to its in-sample results (12.64 out-of-sample RMSE; 5.63 in-sample RMSE). We find that in this case the performance varies substantially across simulation rounds (RMSE SD = 3.4). The lower performance in Los Angeles may be due to the fact that it has fewer disadvantaged neighborhoods (54) compared to New York (79) and London (186). There may also be other unique aspects of gentrification in Los Angeles that limit the usefulness of Airbnb data compared to other cities; for example, gentrification in Los Angeles occurred more recently than in New York City and London and was spurred by external real-estate investment rather than market changes~\citep{clark_LAreport}.

Similar to the in-sample case, the models using all features perform better than the models including only structured or unstructured features, again confirming the importance of unstructured text data. One additional advantage of random forest regression is that it allows us to compute feature importance, i.e., how effective the feature is at reducing uncertainty. Feature importance is calculated using the Mean Decrease Impurity (MDI) score, which represents the percentage of times that a feature is used to split a node weighted by the number of samples it splits. We report the top-5 most important features for the three cities in Table~\ref{table:rf_features}. We find the number of listings and reviews to be the top-2 features for all cities. However, we observe that among the remaining features, many of them are unstructured data features such as Doc2Vec components (the top-3 in terms of correlation with the gentrification score), the LDA Location component, and location words. This again validates the importance of unstructured data and in particular features related to the listings' location.

\section{Discussion and Conclusions}
This work adds to the growing literature employing alternative data sources to predict urban and economic outcomes. We present the first application of Natural Language Processing (NLP) for nowcasting gentrification, and provide evidence that latent information contained in the unstructured text content of Airbnb reviews complements information contained in the structured data; i.e., combining both types of data helps us better explain the process of gentrification. These results have important implications. First, our results highlight the importance of machine learning algorithms for extracting information from unstructured data to create better measures and predictions of socioeconomic indices. Second, NLP tools can find important markers of gentrification from the text content of Airbnb reviews. Third, our work suggests that, while gentrification is generally considered a hidden and slow process~\cite{florida2010great}, Airbnb guests can ``see'' this process and capture it through words. 

Our work does not come without limitations, which may inspire future research. We focus on correlations and predictions, not causality. Estimating causal relationships in our settings is difficult because of simultaneity issues; it could be that Airbnb is more likely to enter in gentrifying neighborhoods, or it could be that Airbnb is speeding up gentrification in the neighborhoods that it enters. To overcome these issues, we would require some exogenous shock that locally affects neighborhoods but not Airbnb. 

In addition, Airbnb data itself has limitations. First, Airbnb data has only grown substantially in recent years (post-2013), which is one of the reasons for why we focus on nowcasting and not forecasting gentrification. Airbnb data availability can also depend on a cities tourism patterns, characteristics, and policies, which means that our approach is far from being universally applicable; for example, it is likely that large and touristy cities like those studied in this work are more suitable for our approach. We further caution that models solely based on Airbnb data will skew toward the perspective of its user population, which is predominantly affluent, educated, and younger (similar demographics to in-movers in the gentrification process). Policymakers should weigh the insights and biases of alternative data sources like Airbnb when using such data to inform decisions.


As a potential way to reduce the biases associated with relying on data from a single platform, future work could leverage an ensemble of user-generated data from different platforms. As prior work suggests~\citep{glaeser_yelp,naik_streetscore}, there are plenty of user-generated data that could be used to improve our predictions and create more accurate models, including data obtained from Yelp or Google. Other unstructured data besides text could be used as well, such as photos or videos. Moreover, in the future, the availability of additional historical data from Airbnb and other platforms may be used to forecast (instead of nowcast) gentrification. While forecasting may be more beneficial for policymakers, it may also be more challenging than nowcasting. Therefore, future research may provide more insights about forecasting gentrification and the challenges that need to be addressed to achieve good results. 



The availability of big data and machine learning tools to analyze user-generated information is becoming increasingly useful in many settings, from marketing~\citep{liu2019large} to economics~\citep{glaeser_streetscore} to urban social science~\citep{smith2013finger}. In the context of this paper, urban social science, such data availability is rapidly changing and improving how cities and municipalities measure important outcomes which, in turn, will directly improve policymaking, and, more generally, our understanding of how cities change and evolve.

\begin{acks}
The authors thank Isaac Gelman, Lauren Phillips, Nat Redfern, and Sahil Agarwal for their assistance with this research. We also acknowledge the USC Center for AI in Society’s Student Branch for providing us with computing resources.
\end{acks}

\section*{Online Resources}
Data and code for this work are available at: \url{https://github.com/shomikj/airbnb_gentrification}.

\bibliographystyle{ACM-Reference-Format}
\bibliography{bibliography}

\newpage
\appendix

\section{Appendix}

\vspace{0.25cm}
\begin{table}[ht] \centering \mytablesize
\caption{Regression for Gentrification Score with Race.}
\begin{tabular}{
p{2.5cm}
p{2cm}
S[table-number-alignment = center, group-digits=true, table-format= 2.2, group-separator={}, round-mode=places, round-precision=2]
S[table-number-alignment = center, group-digits=true, table-format= 2.2, group-separator={}, round-mode=places, round-precision=2]
S[table-number-alignment = center, group-digits=true, table-format= 2.2, group-separator={}, round-mode=places, round-precision=2]
S[table-number-alignment = center, group-digits=true, table-format= 2.2, group-separator={}, round-mode=places, round-precision=2]
} 
\toprule
\multirow{2}{*}{Model} & \multirow{2}{*}{City} & \multicolumn{4}{c}{Root Mean Squared Error} \\ 
\cmidrule(lr){3-6}
& & \multicolumn{1}{c}{Baseline} &
\multicolumn{1}{c}{Structured} & \multicolumn{1}{c}{Unstructured} &
\multicolumn{1}{c}{All Features}\\
\midrule
In-Sample & New York & 13.223 & 8.15 & 5.582 & 5.128\\
Linear & Los Angeles & 13.778 & 10.409 & 5.77 & 4.565\\
Regression & London & \multicolumn{4}{c}{--------------------------------- N/A ---------------------------------}\\
\addlinespace
\midrule
Out-of-Sample & New York & 13.186 & 8.119 & 8.755 & 8.064\\
Random Forest & Los Angeles & 13.705 & 12.303 & 11.474 & 11.498\\
Regression & London & \multicolumn{4}{c}{--------------------------------- N/A ---------------------------------}\\
\addlinespace
\bottomrule
\multicolumn{6}{l}{\textit{Note: Random Forest results averaged over 100 iterations with 50\%-50\% train-test split.}} \\
\end{tabular}
\label{table:regA1}
\end{table}

\vspace{0.25cm}
\begin{table}[ht] \centering \mytablesize
\caption{Regression for Race.}
\begin{tabular}{
p{2.5cm}
p{2cm}
S[table-number-alignment = center, group-digits=true, table-format= 2.2, group-separator={}, round-mode=places, round-precision=2]
S[table-number-alignment = center, group-digits=true, table-format= 2.2, group-separator={}, round-mode=places, round-precision=2]
S[table-number-alignment = center, group-digits=true, table-format= 2.2, group-separator={}, round-mode=places, round-precision=2]
S[table-number-alignment = center, group-digits=true, table-format= 2.2, group-separator={}, round-mode=places, round-precision=2]
} 
\toprule
\multirow{2}{*}{Model} & \multirow{2}{*}{City} & \multicolumn{4}{c}{Root Mean Squared Error} \\ 
\cmidrule(lr){3-6}
& & \multicolumn{1}{c}{Baseline} &
\multicolumn{1}{c}{Structured} & \multicolumn{1}{c}{Unstructured} &
\multicolumn{1}{c}{All Features}\\
\midrule
In-Sample & New York & 9.02 & 6.132 & 4.165 & 3.79 \\
Linear & Los Angeles & 9.954 & 5.515 & 2.969 & 2.431 \\
Regression & London & \multicolumn{4}{c}{--------------------------------- N/A ---------------------------------}\\
\addlinespace
\midrule
Out-of-Sample & New York & 9.051 & 5.693 & 6.493 & 5.816 \\
Random Forest & Los Angeles & 9.923 & 6.466 & 6.322 & 6.106 \\
Regression & London & \multicolumn{4}{c}{--------------------------------- N/A ---------------------------------}\\
\addlinespace
\bottomrule
\multicolumn{6}{l}{\textit{Note: Random Forest results averaged over 100 iterations with 50\%-50\% train-test split.}} \\
\end{tabular}
\label{table:regA1}
\end{table}

\begin{table}[ht] \centering \mytablesize
\caption{Regression for Age.}
\begin{tabular}{
p{2.5cm}
p{2cm}
S[table-number-alignment = center, group-digits=true, table-format= 2.2, group-separator={}, round-mode=places, round-precision=2]
S[table-number-alignment = center, group-digits=true, table-format= 2.2, group-separator={}, round-mode=places, round-precision=2]
S[table-number-alignment = center, group-digits=true, table-format= 2.2, group-separator={}, round-mode=places, round-precision=2]
S[table-number-alignment = center, group-digits=true, table-format= 2.2, group-separator={}, round-mode=places, round-precision=2]
} 
\toprule
\multirow{2}{*}{Model} & \multirow{2}{*}{City} & \multicolumn{4}{c}{Root Mean Squared Error} \\ 
\cmidrule(lr){3-6}
& & \multicolumn{1}{c}{Baseline} &
\multicolumn{1}{c}{Structured} & \multicolumn{1}{c}{Unstructured} &
\multicolumn{1}{c}{All Features}\\
\midrule
In-Sample & New York & 16.497 & 11.311 & 9.236 & 8.038\\
Linear & Los Angeles & 14.768 & 13.094 & 9.733 & 8.616\\
Regression & London & 14.05 & 12.827 & 11.417 & 11.212\\
\addlinespace
\midrule
Out-of-Sample & New York &  16.545 & 12.51 & 14.148 & 13.35\\
Random Forest & Los Angeles &  14.532 & 15.952 & 15.771 & 15.713 \\
Regression & London & 14.023 & 13.616 & 13.101 & 13.161 \\
\addlinespace
\bottomrule
\multicolumn{6}{l}{\textit{Note: Random Forest results averaged over 100 iterations with 50\%-50\% train-test split.}} \\
\end{tabular}
\label{table:regA1}
\end{table}

\vspace{0.25cm}
\begin{table}[ht] \centering \mytablesize
\caption{Regression for Education.}
\begin{tabular}{
p{2.5cm}
p{2cm}
S[table-number-alignment = center, group-digits=true, table-format= 2.2, group-separator={}, round-mode=places, round-precision=2]
S[table-number-alignment = center, group-digits=true, table-format= 2.2, group-separator={}, round-mode=places, round-precision=2]
S[table-number-alignment = center, group-digits=true, table-format= 2.2, group-separator={}, round-mode=places, round-precision=2]
S[table-number-alignment = center, group-digits=true, table-format= 2.2, group-separator={}, round-mode=places, round-precision=2]
} 
\toprule
\multirow{2}{*}{Model} & \multirow{2}{*}{City} & \multicolumn{4}{c}{Root Mean Squared Error} \\ 
\cmidrule(lr){3-6}
& & \multicolumn{1}{c}{Baseline} &
\multicolumn{1}{c}{Structured} & \multicolumn{1}{c}{Unstructured} &
\multicolumn{1}{c}{All Features}\\
\midrule
In-Sample & New York & 12.882 & 7.926 & 5.726 & 5.163\\
Linear & Los Angeles & 13.834 & 10.934 & 4.878 & 4.036\\
Regression & London & 20.999 & 18.226 & 16.608 & 15.897\\
\addlinespace
\midrule
Out-of-Sample & New York &  13.077 & 7.135 & 8.423 & 7.369\\
Random Forest & Los Angeles &  12.731 & 11.887 & 10.643 & 10.584\\
Regression & London & 21.009 & 18.817 & 18.811 & 18.531\\
\addlinespace
\bottomrule
\multicolumn{6}{l}{\textit{Note: Random Forest results averaged over 100 iterations with 50\%-50\% train-test split.}} \\
\end{tabular}
\label{table:regA1}
\end{table}

\vspace{0.25cm}
\begin{table}[ht] \centering \mytablesize
\caption{Regression for Income.}
\begin{tabular}{
p{2.5cm}
p{2cm}
S[table-number-alignment = center, group-digits=true, table-format= 2.2, group-separator={}, round-mode=places, round-precision=2]
S[table-number-alignment = center, group-digits=true, table-format= 2.2, group-separator={}, round-mode=places, round-precision=2]
S[table-number-alignment = center, group-digits=true, table-format= 2.2, group-separator={}, round-mode=places, round-precision=2]
S[table-number-alignment = center, group-digits=true, table-format= 2.2, group-separator={}, round-mode=places, round-precision=2]
} 
\toprule
\multirow{2}{*}{Model} & \multirow{2}{*}{City} & \multicolumn{4}{c}{Root Mean Squared Error} \\ 
\cmidrule(lr){3-6}
& & \multicolumn{1}{c}{Baseline} &
\multicolumn{1}{c}{Structured} & \multicolumn{1}{c}{Unstructured} &
\multicolumn{1}{c}{All Features}\\
\midrule
In-Sample & New York & 9.026 & 6.451 & 4.695 & 3.914\\
Linear & Los Angeles & 10.779 & 9.03 & 4.775 & 4.215\\
Regression & London & 13.861 & 12.77 & 10.93 & 10.735\\
\addlinespace
\midrule
Out-of-Sample & New York & 8.856 & 7.411 & 7.816 & 7.635 \\
Random Forest & Los Angeles & 9.578 & 9.884 & 8.932 & 8.976 \\
Regression & London & 13.783 & 13.789 & 12.853 & 12.927\\
\addlinespace
\bottomrule
\multicolumn{6}{l}{\textit{Note: Random Forest results averaged over 100 iterations with 50\%-50\% train-test split.}} \\
\end{tabular}
\label{table:regA1}
\end{table}

\vspace{0.25cm}
\begin{table}[ht] \centering \mytablesize
\caption{Regression for Rent.}
\begin{tabular}{
p{2.5cm}
p{2cm}
S[table-number-alignment = center, group-digits=true, table-format= 2.2, group-separator={}, round-mode=places, round-precision=2]
S[table-number-alignment = center, group-digits=true, table-format= 2.2, group-separator={}, round-mode=places, round-precision=2]
S[table-number-alignment = center, group-digits=true, table-format= 2.2, group-separator={}, round-mode=places, round-precision=2]
S[table-number-alignment = center, group-digits=true, table-format= 2.2, group-separator={}, round-mode=places, round-precision=2]
} 
\toprule
\multirow{2}{*}{Model} & \multirow{2}{*}{City} & \multicolumn{4}{c}{Root Mean Squared Error} \\ 
\cmidrule(lr){3-6}
& & \multicolumn{1}{c}{Baseline} &
\multicolumn{1}{c}{Structured} & \multicolumn{1}{c}{Unstructured} &
\multicolumn{1}{c}{All Features}\\
\midrule
In-Sample & New York & 10.974 & 6.918 & 5.768 & 4.539\\
Linear & Los Angeles & 15.039 & 12.52 & 7.581 & 7.019\\
Regression & London & 23.273 & 19.252 & 17.137 & 16.627\\
\addlinespace
\midrule
Out-of-Sample & New York & 11.078 & 9.333 & 9.335 & 9.184 \\
Random Forest & Los Angeles &  14.642 & 15.317 & 13.781 & 14.089 \\
Regression & London & 23.238 & 20.504 & 20.663 & 19.866 \\
\addlinespace
\bottomrule
\multicolumn{6}{l}{\textit{Note: Random Forest results averaged over 100 iterations with 50\%-50\% train-test split.}} \\
\end{tabular}
\label{table:regA1}
\end{table}

\begin{figure}[h!]
\centering
\subfloat[][New York]{\includegraphics[width=0.45\linewidth]{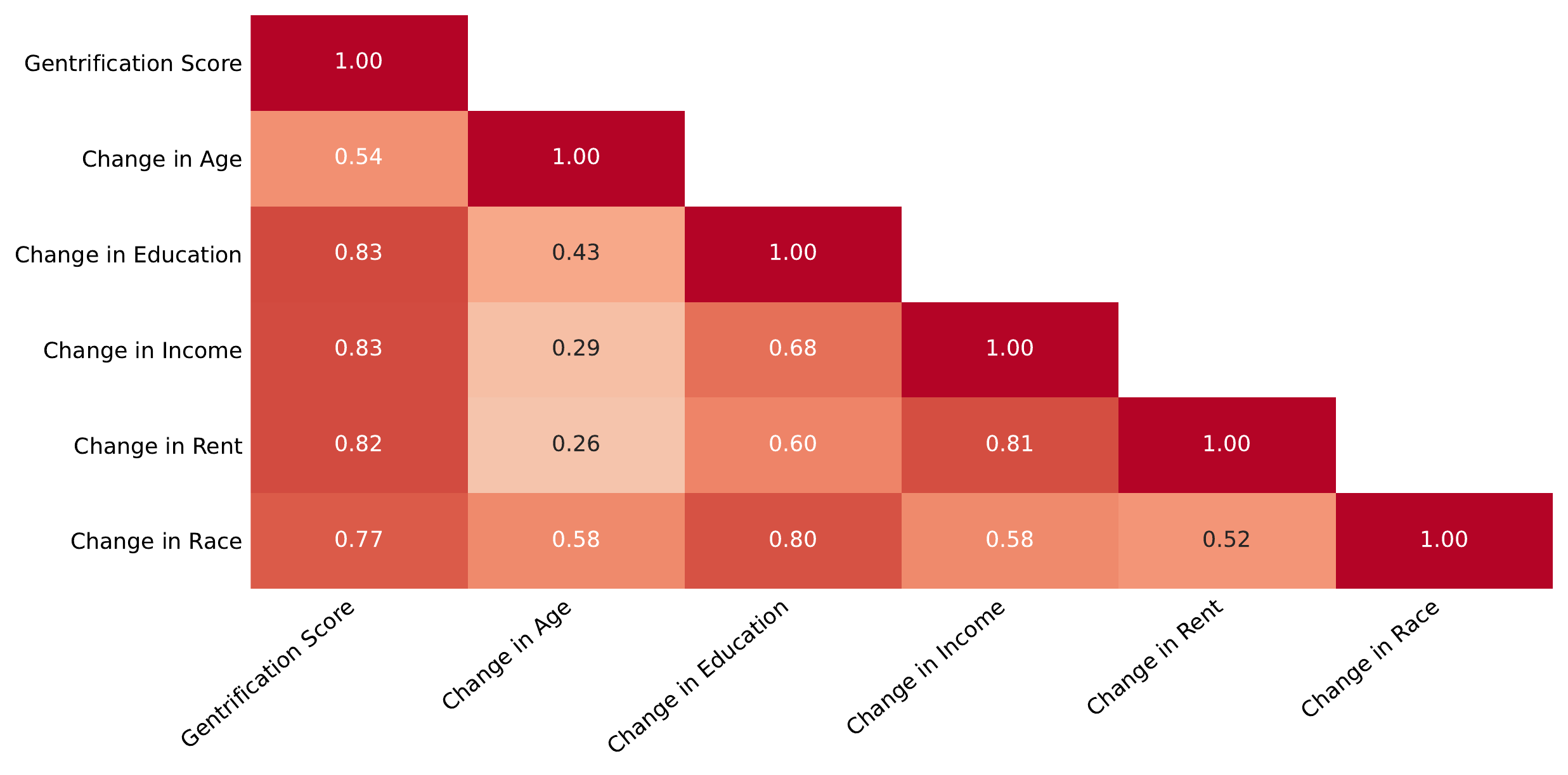}}
\subfloat[][Los Angeles]{\includegraphics[width=0.45\linewidth]{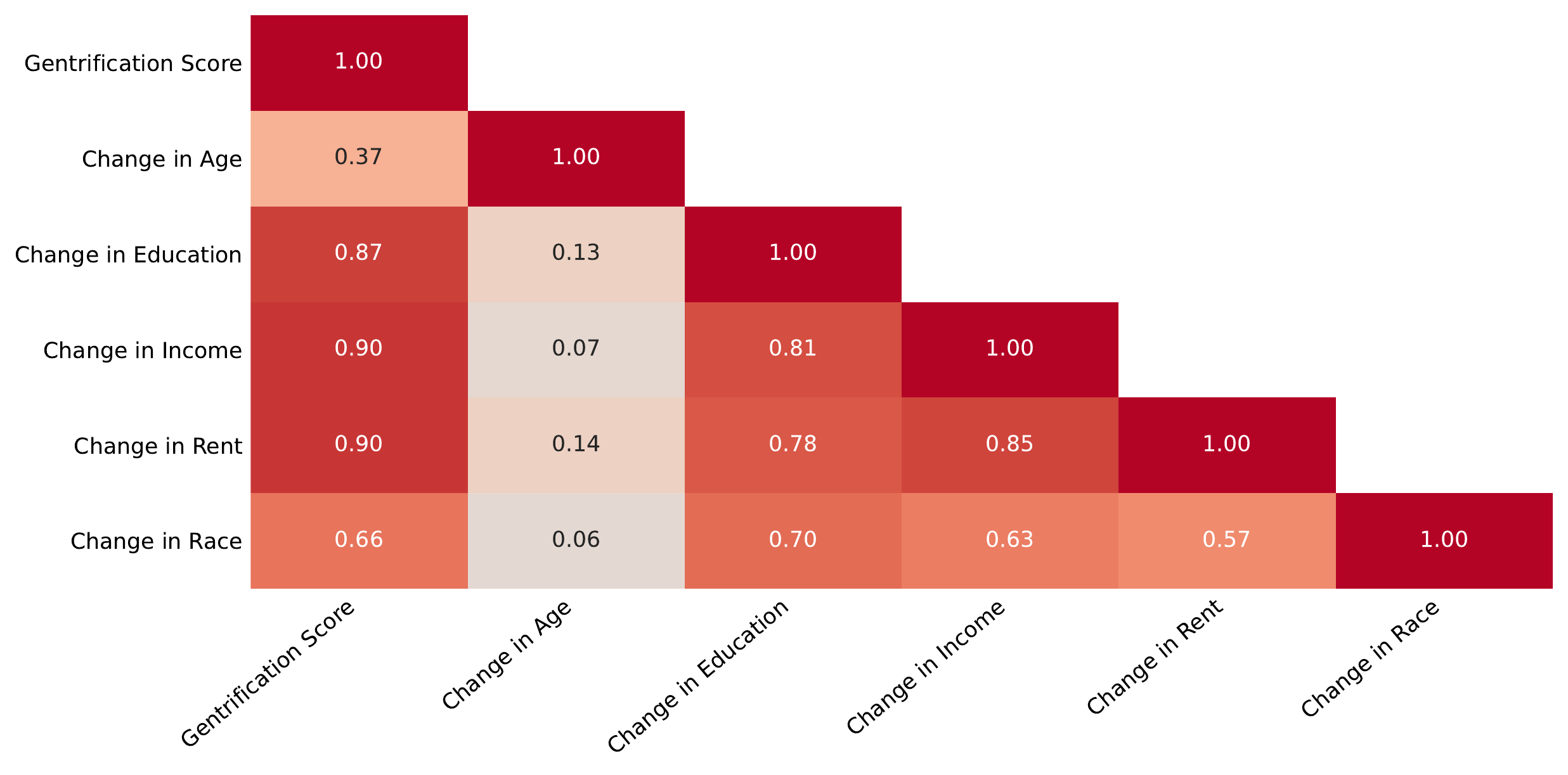}}
\newline
\subfloat[][London]{\includegraphics[width=0.45\linewidth]{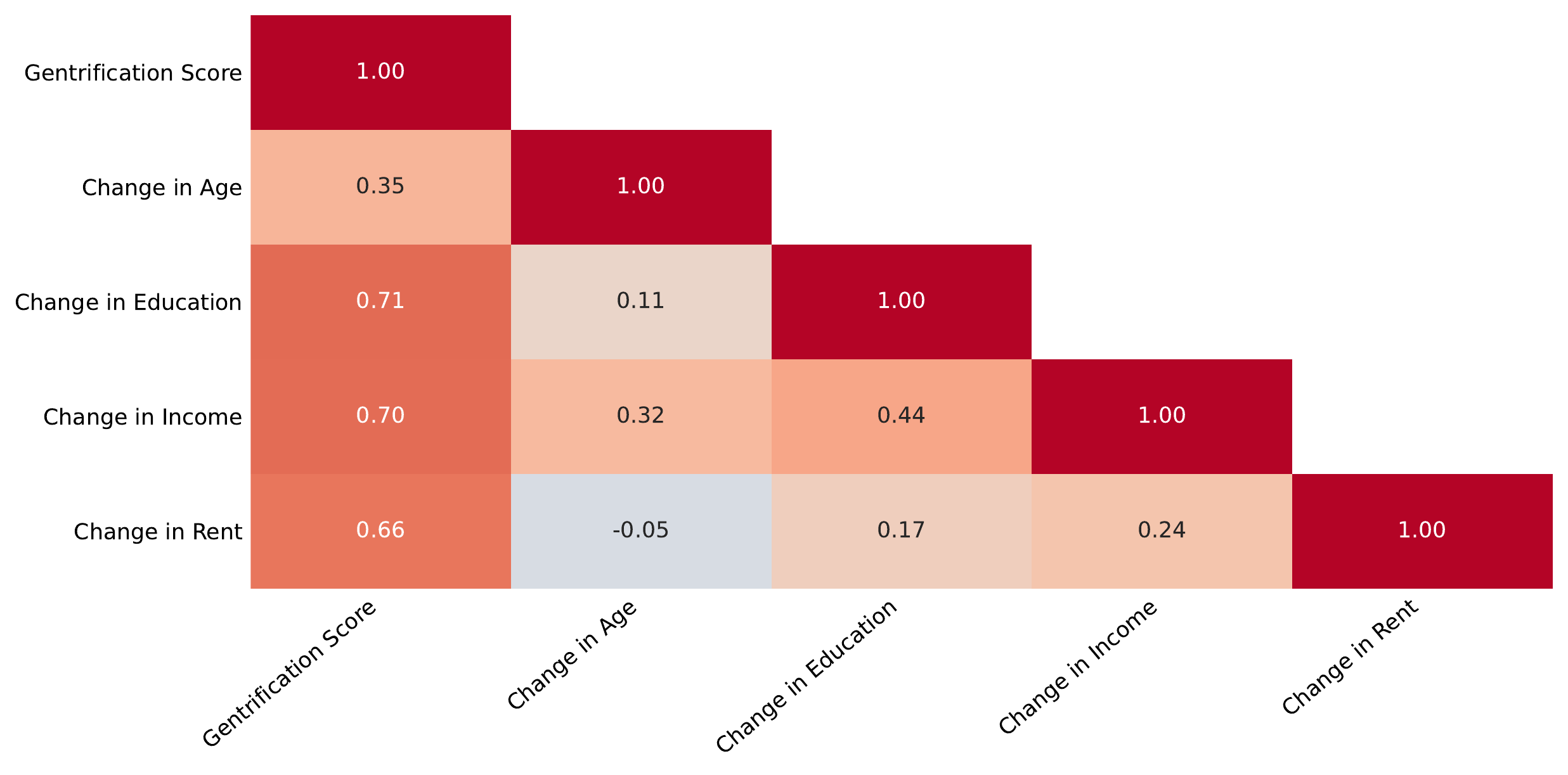}}
\caption{Cross-Correlation Among Gentrification Score and Socioeconomic Variables.}
\label{fig:socioeconomic_correlation}
\end{figure}

\begin{figure}[h!]
\centering
\subfloat[][New York]{\includegraphics[width=0.75\linewidth]{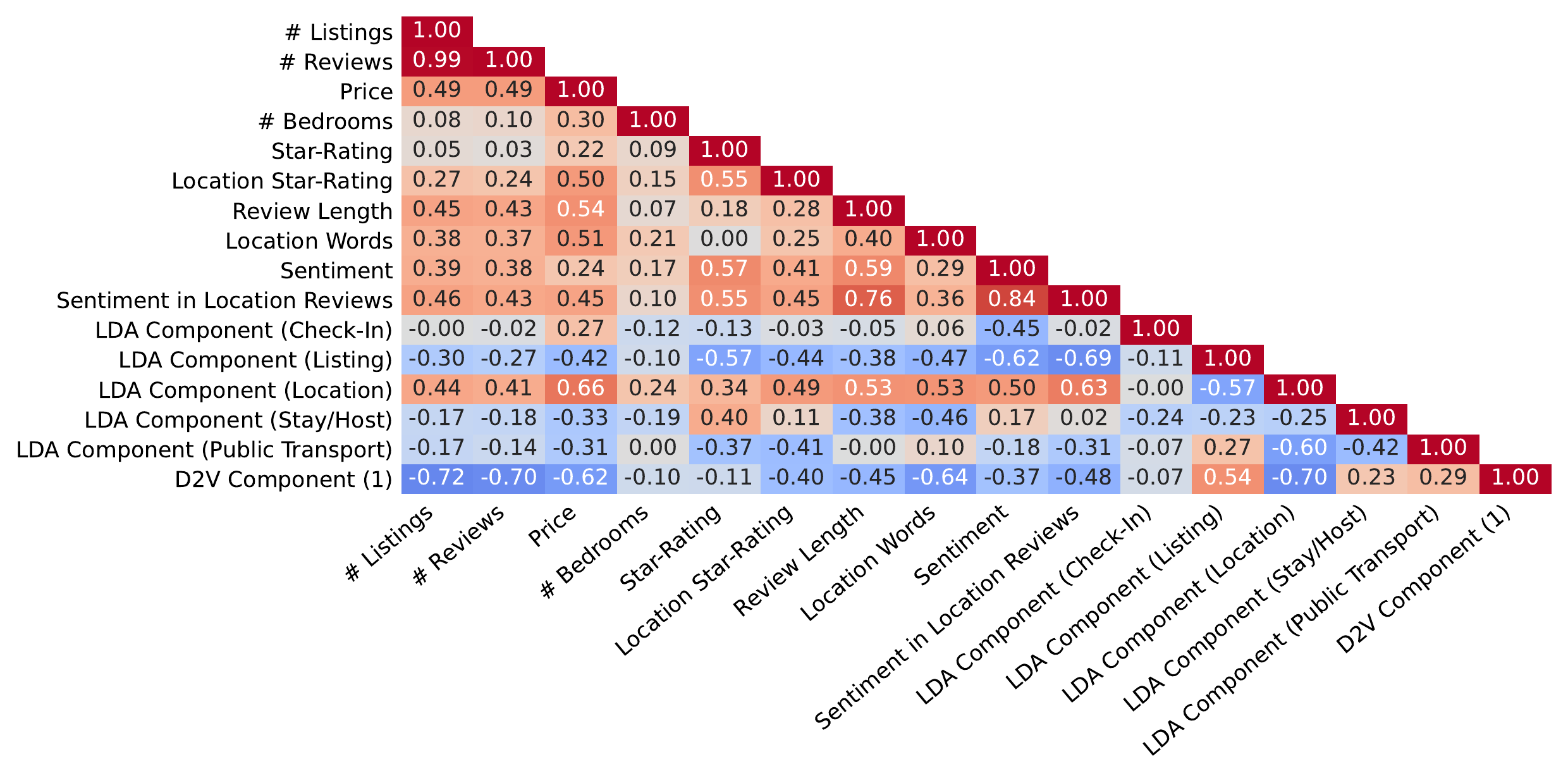}}
\newline
\subfloat[][Los Angeles]{\includegraphics[width=0.75\linewidth]{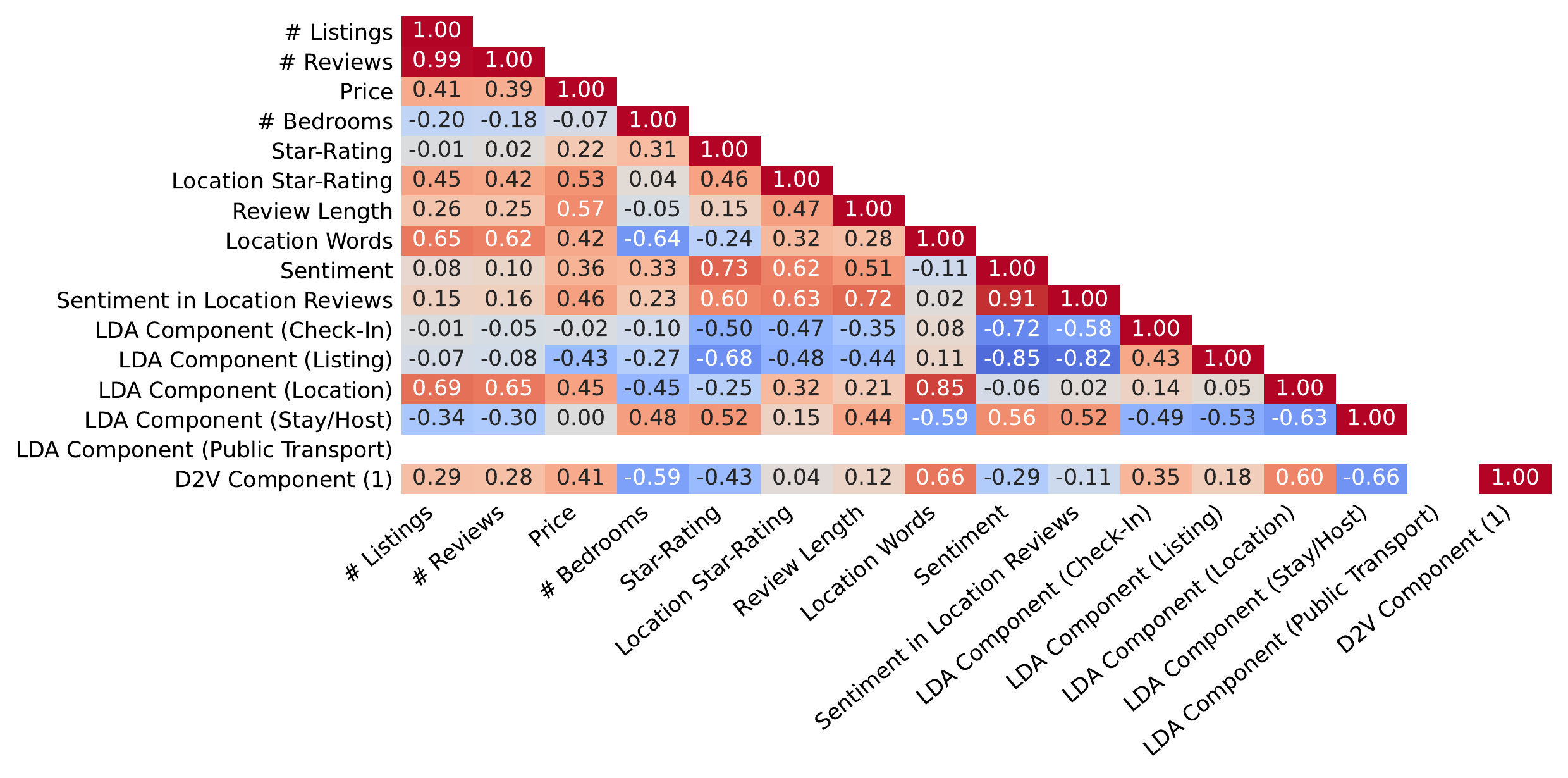}}
\newline
\subfloat[][London]{\includegraphics[width=0.75\linewidth]{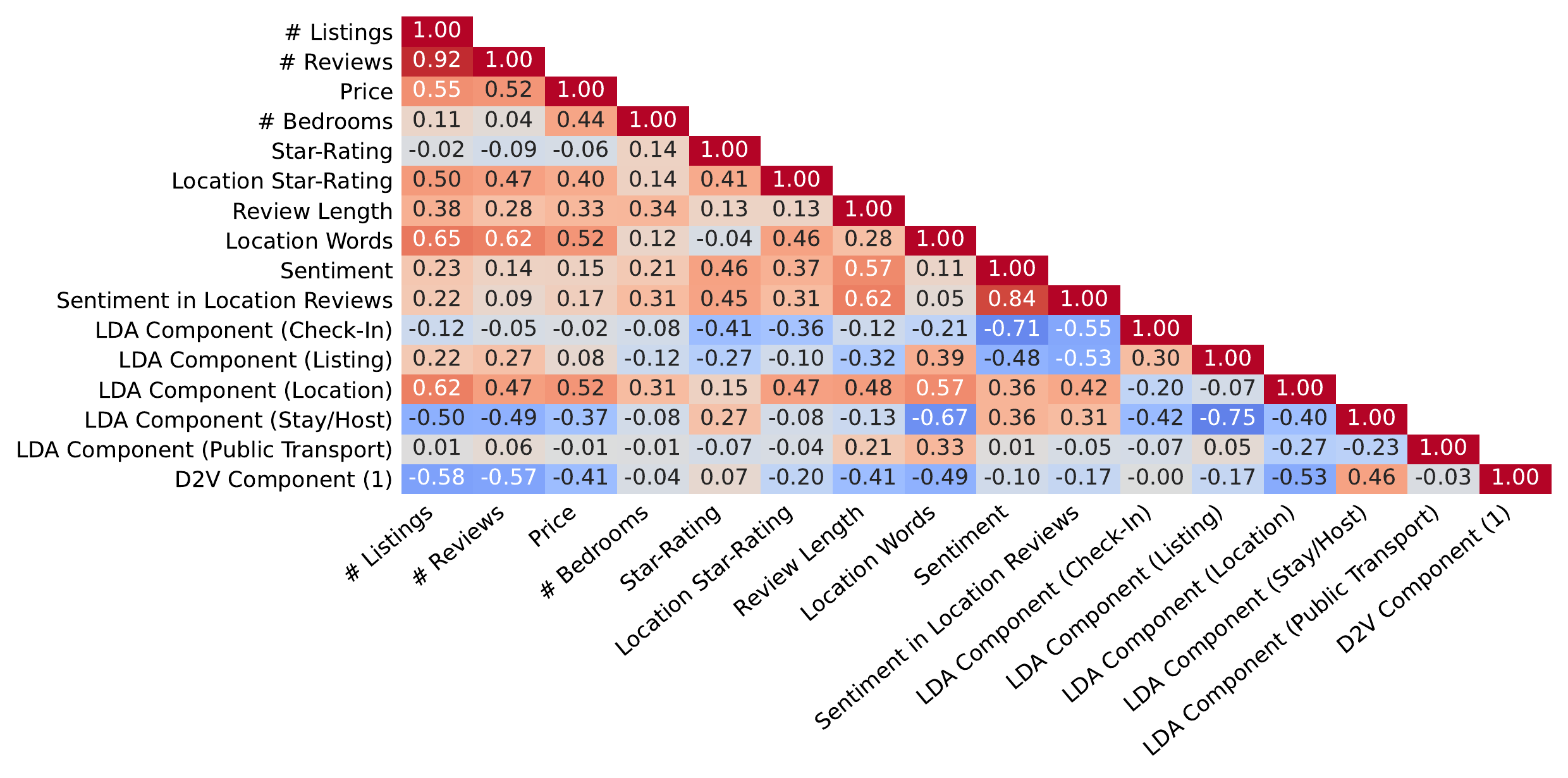}}
\caption{Cross-Correlation Among Airbnb Features.}
\label{fig:inter_correlations}
\end{figure}


\end{document}